\documentclass[a4paper,fleqn,usenatbib]{mnras}

\usepackage{newtxtext,newtxmath}

\usepackage[T1]{fontenc}
\usepackage{ae,aecompl}

\usepackage{graphicx}	
\usepackage{amsmath}	
\usepackage{amssymb}	
\usepackage{epstopdf}	
\usepackage{soul}
\usepackage{pifont}
\newcommand{\xmark}{\ding{55}}%
\newcommand{\Mst}{M_*}%
\title[Magnetic braking in galaxies]{How primordial magnetic fields shrink galaxies}
\author[S. Martin-Alvarez et al.]{Sergio Martin-Alvarez,$^{1}$\thanks{E-mail: smartin@ast.cam.ac.uk (SMA)}
Adrianne Slyz,$^{2}$ Julien Devriendt$^{2,3}$ \newauthor
and Carlos G\'omez-Guijarro$^{4}$ \\
$^{1}$Institute of Astronomy and Kavli Institute for Cosmology, University of Cambridge, Madingley Road, Cambridge CB3 0HA, UK\\
$^{2}$Subdepartment of Astrophysics, University of Oxford, Keble Road, Oxford, OX1 3RH, UK\\
$^{3}$Universit\'e de Lyon, Universit\'e Lyon 1, ENS de Lyon, CNRS, Centre de Recherche Astrophysique de Lyon, UMR 5574, F-69230 Saint-Genis-Laval, France\\
$^{4}$ AIM, CEA, CNRS, Université Paris-Saclay, Université Paris Diderot, Sorbonne Paris Cité, F-91191 Gif-sur-Yvette, France\\
}

\date{Accepted XXX. Received YYY; in original form ZZZ}

\pubyear{2020}
\begin{document}

\definecolor{B20}{rgb}{0.930951, 0.841744, 0.916397}
\definecolor{B14}{rgb}{0.83004, 0.650319, 0.797752}
\definecolor{B13}{rgb}{0.685202, 0.453527, 0.643476}
\definecolor{B12}{rgb}{0.558209, 0.327623, 0.517503}
\definecolor{B11}{rgb}{0.366272, 0.222226, 0.343123}
\definecolor{B10}{rgb}{0.206948, 0.148732, 0.197668}

\label{firstpage}
\pagerange{\pageref{firstpage}--\pageref{lastpage}}
\maketitle

\begin{abstract}
As one of the prime contributors to the interstellar medium energy budget, magnetic fields naturally play a part in shaping the evolution of galaxies. Galactic magnetic fields can originate from strong primordial magnetic fields provided these latter remain below current observational upper limits. To understand how such magnetic fields would affect the global morphological and dynamical properties of galaxies, we use a suite of high-resolution constrained transport magneto-hydrodynamic cosmological zoom simulations where we vary the initial magnetic field strength and configuration along with the prescription for stellar feedback. We find that strong primordial magnetic fields delay the onset of star formation and drain the rotational support of the galaxy, diminishing the radial size of the galactic disk and driving a higher amount of gas towards the centre. This is also reflected in mock $UVJ$ observations by an increase in the light profile concentration of the galaxy. We explore the possible mechanisms behind such a reduction in angular momentum, focusing on magnetic braking. Finally, noticing that the effects of primordial magnetic fields are amplified in the presence of stellar feedback, we briefly discuss whether the changes we measure would also be expected for galactic magnetic fields of non-primordial origin. 
\end{abstract}


\section{Introduction}
While there is consensus that magnetic fields should pervade our Universe on cosmological scales, not much is known about the primordial magnetogenesis scenario which generates them. Depending on specifics, this can lead to substantially different distributions for primordial magnetic field configurations, varying in normalization, coherence length, and/or spectral index. Our understanding of the cosmic magnetic field present-day properties is not in significantly better shape. Observational constraints on its strength span several orders of magnitude. The upper limit from CMB anisotropies yields $B_0 < 4 \cdot 10^{-9}$ G \citep{Planck16}. This value is reduced to $B_0 < 1 \cdot 10^{-9}$ G when combined with results from SPT \citep{Pogosian18}. Ultra-high energy cosmic rays (UHECRs) provide an alternative upper limit $B_0 < 0.5 \cdot 10^{-9}$ G \citep{Bray18}, of a similar order of magnitude to that from Planck. A lower limit can be derived from the broadening of $\gamma$-ray emission from distant TeV Blazars \citep{Neronov10, Taylor11}. This yields $B_0 \gtrsim 10^{-16}$ G\footnote{Note however that these results have alternative interpretations which dispute the indicated value \citep[e.g.][]{Broderick12}.}.

Improving our understanding of cosmic magnetism is an important task: magnetic fields frequently play an important role, on scales ranging from individual star forming regions to the largest virialised structures in our Universe. They affect the properties of radio haloes \citep{Marinacci17}, galaxy clusters and AGN \citep{Yang16, Egan16}. Furthermore, strong primordial magnetic fields can influence structure formation, acting as a reheating source during recombination \citep{Trivedi18}, modifying density and vorticity perturbations \citep{Tsagas00}, dark matter haloes \citep{Varalakshmi17, Cheera18} or baryonic density perturbations \citep{Kim96}. The configuration of primordial magnetic fields needs to be taken into account for the propagation of ultra-high energy cosmic rays through cosmic voids \citep{Wittor17, Alves-Batista17}, and could affect the magnetisation of the filaments in the cosmic web \citep{Marinacci15}.

For these reasons, the properties and evolution of primordial magnetic fields in a cosmological context have been studied by various authors \citep{Vazza14, Marinacci15, Gheller16}. However, regardless of all these potential effects, the direct detection of magnetic fields, as yet, has been limited to individual galaxies and galaxy clusters, with perhaps the exception of some radio ridges \citep{Govoni2019}.

Even in galaxies, the origin of the observed magnetic fields (typically of several $\mu$G in strength, \citealt{Beck15}) remains unclear. The most direct explanation is a primordial origin, as a result of an energetic magnetogenesis process potentially followed by a modest amount of amplification. Alternatively, galactic magnetic fields might have evolved through in-situ large amplification of weak primordial seeds \citep{Pakmor14, Vazza14, Martin-Alvarez18}; or been seeded through processes such as stellar winds, supernovae (SNe), or AGN \citep{Beck13a, Butsky17, Vazza17, KMA18} acting as small-scale batteries. Other possibilities are their generation during reionization \citep{Durrive17}, or their amplification prior to accretion onto galaxies or galaxy clusters by shock-induced turbulence \citep{Kulsrud97, Suoqing16}, potentially tracing the spectrum of cosmic shock waves \citep{Martin-Alvarez17}. Amongst these diverse scenarios, the main advantage provided by a primordial origin is a simultaneous explanation of the existence of cosmic, cluster, and galactic magnetic fields.

Determining how magnetic fields reach their current state is of particular importance for galaxies. Galactic haloes are expected to erase any memory of weak primordial configurations, rendering the magnetic seeding scenario irrelevant. However, whenever strong enough primordial magnetic fields are considered, the amount of magnetic energy contained in the pristine gas becomes non-negligible, both inside and around galaxies. As a consequence, strong primordial magnetic fields ($B_0 \gtrsim 10^{-10} G$) are expected to impact some properties of galaxies \citep{Dubois10, Marinacci16, Safarzadeh19}.

While current simulations are able to generate realistic galaxies by resolving the multi-phase structure of the ISM and account for a variety of important physical processes \citep[e.g][]{Hopkins14, Kimm15, Grisdale17}; they generally neglect magnetic fields, despite magnetic energy being in equipartition with turbulent energy and comparable or above thermal energy, even at high redshifts \citep[$z \lesssim 2$,][]{Bernet08, Wolfe08, Mulcahy14, Mao2017, Mulcahy17}. Accordingly, magnetic fields should affect the ISM multi-phase structure and dynamics \citep{Moss07, Villagran17, Xu18, Kortgen2019}. They could destabilise spiral arms \citep{Inoue18} or reduce the number of star formation sites in the ISM \citep{Hennebelle14}. Magnetic fields have also been proposed as a mechanism to reduce angular momentum in galaxies \citep{Sparke82}, or drive inward gas flows in galactic bars \citep{Moss00, Beck05}. These latter processes could significantly decrease the size of a galaxy, alter its morphology, and contribute to the formation of galactic bulges. Last but not least, magnetic fields are also known to be relevant on smaller sub-galactic scales, being one of the key actors regulating star formation within molecular clouds \citep[e.g.][]{Tan04, McKee07, Hull17}.

In this work, we study the impact that primordial magnetic fields have on the formation of an individual Milky Way-like galaxy in a cosmological context, re-simulated at high-resolution under various simple primordial magnetic configurations. We quantify how these primordial magnetic fields affect the global morphological and dynamical properties of galaxies at high redshift ($z \geq 2$), and assess the robustness of their influence employing different initial strengths for the primordial magnetic fields and vis-\`a-vis stellar feedback. We also explore how these effects translate into observational signatures at $z = 2$ by the use of mock $UVJ$ observations. We argue some of the changes are caused by magnetic braking and discuss whether they would still happen had galactic magnetic fields had a non-primordial origin.

In Section \ref{s:NumericalMethods} we introduce the numerical method employed and the suite of simulations used in this work. We present our main results in Section \ref{s:Results}, where we first address the impact of primordial magnetic fields on the galaxy (Section \ref{ss:GalacticPhysics}). We end this section by describing magnetic braking. Section \ref{ss:LowZ} is devoted to observational signatures at $z = 2$. We summarise our main conclusions in Section \ref{s:Conclusions}.

\section{Numerical methods}
\label{s:NumericalMethods}

\begin{figure*}%
        \includegraphics[width=1.9\columnwidth]{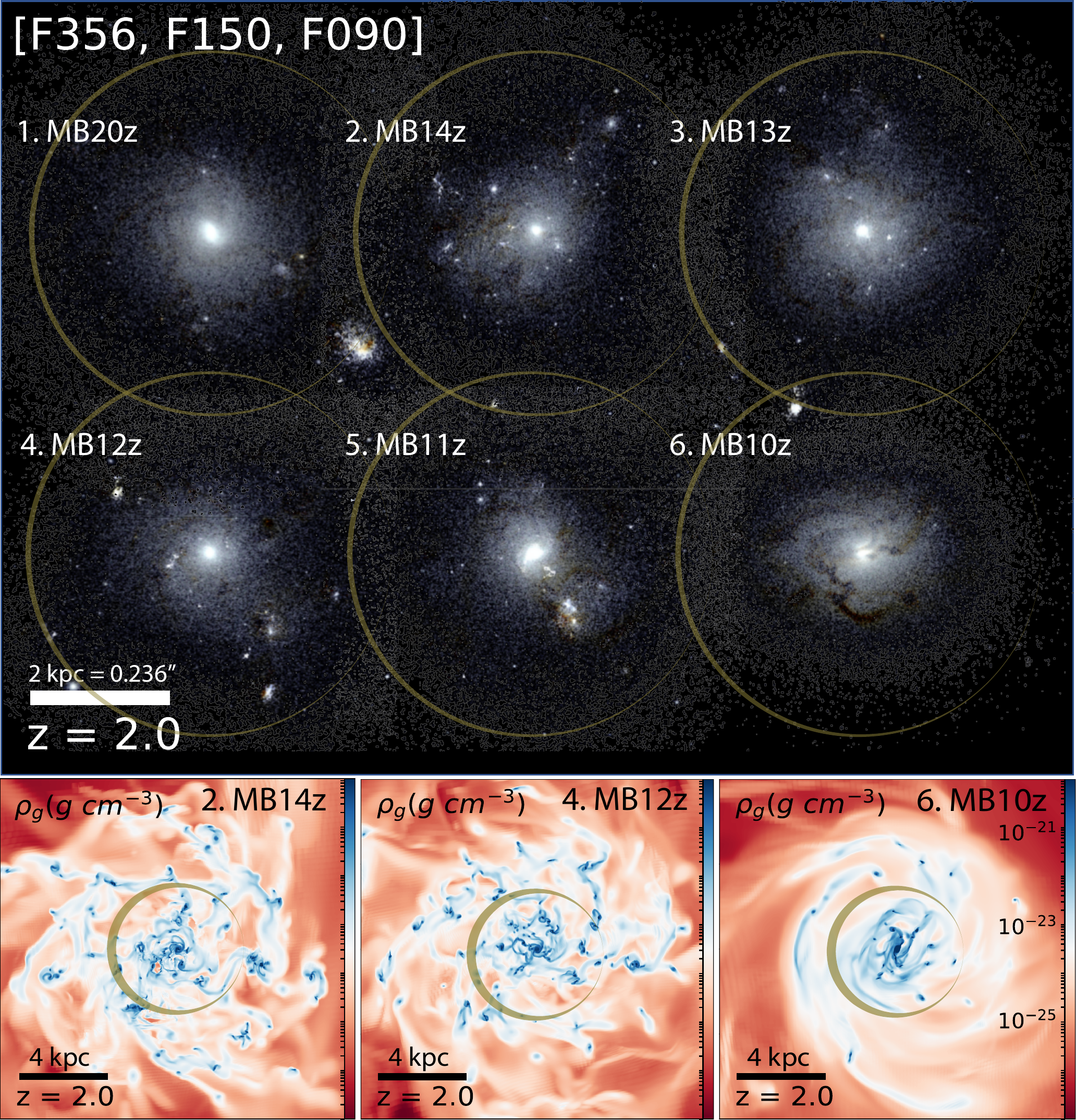}%
		\caption{(Top) Rest-frame $UVJ$ mock observations through the [$F090W$, $F150W$, $F356W$] {\it JWST}/NIRCam filters of the {\sc nut} galaxy at $z = 2$. The strength of the magnetic field $B_0$ increases by ten orders of magnitude from {\it 1.MB20z} ($B_0 = 3 \cdot 10^{-20}$ G) to {\it 6.MB10z} ($B_0 = 3 \cdot 10^{-10}$ G; see text for details). Each panel is centred on the galaxy. On both the mock images and the gas density projections, we include golden circles of radius 3 physical kpc, centred on each galaxy to aid the visual comparison of their dimensions (see also the scale bar on the Figure). (Bottom) Gas density projections $\rho_g$ of some of these galaxies, in panels 16 kpc across.}%
\label{GalaxiesZ2}%
\end{figure*}

Our set of numerical magneto-hydrodynamical (MHD) simulations are generated with our own modified version of the publicly available code {\sc ramses} \citep{Teyssier02, Teyssier06, Fromang06}. {\sc ramses} couples a tree-based adaptive mesh refinement (AMR) Eulerian treatment of the gas with an N-body solver for the dark matter and stellar components. In this section, we briefly introduce the simulation setup, the seeding employed for the magnetic component, and the suite of simulations performed. We also explain the procedures used to make the main measurements and to generate mock observations of the simulated galaxies presented in this paper (shown in Fig. \ref{GalaxiesZ2}).

\subsection{Numerical setup}
\label{ss:Setup}

\begin{figure}%
		\includegraphics[width=\columnwidth]{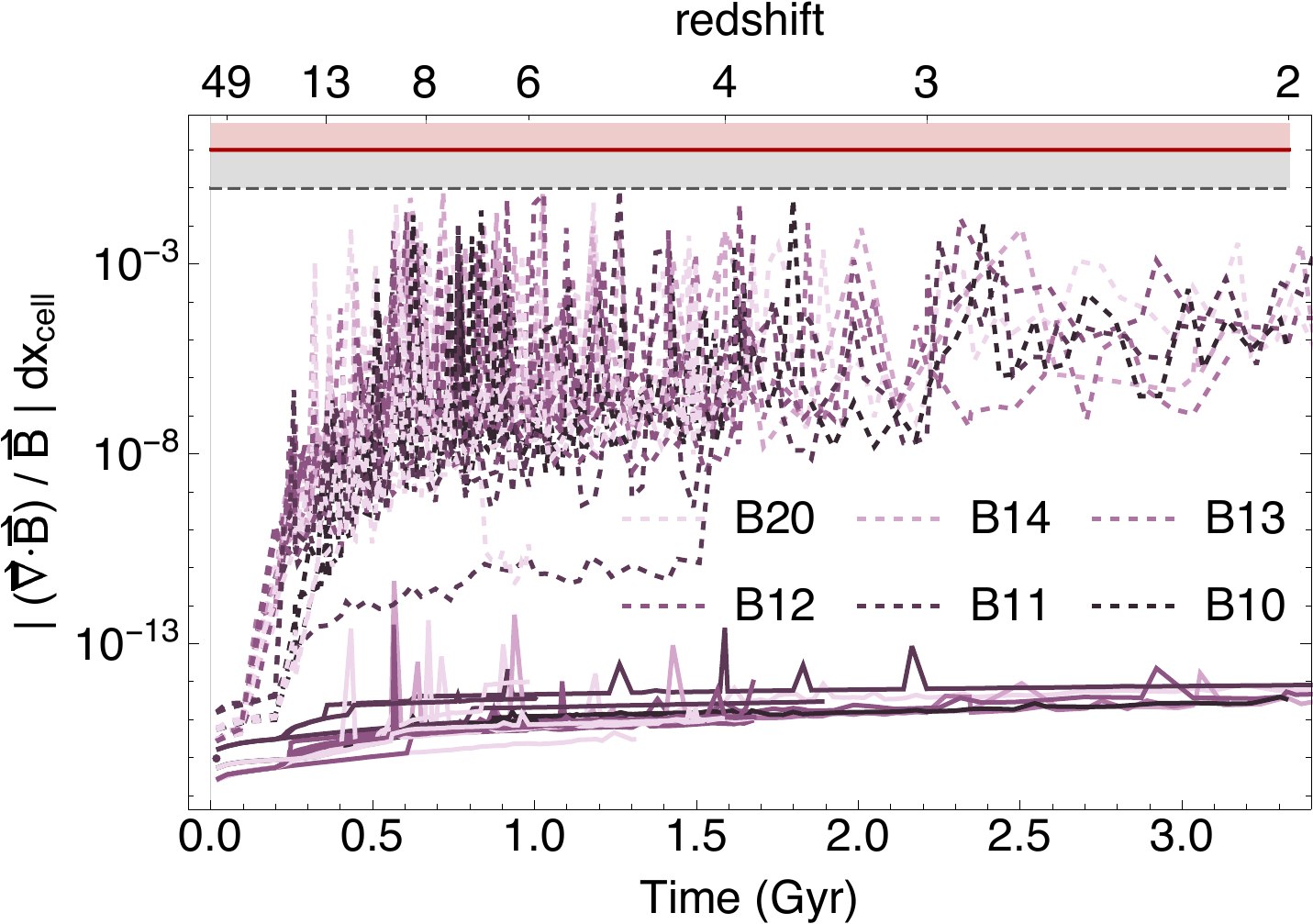}%
		\caption{Magnetic field divergence $|\vec{\nabla} \cdot \vec{B}|$ to total magnetic field $|\vec{B}|$ per cell length $\Delta\text{x}_\text{cell}$ ratio in the entire numerical domain for all simulations (see Table \ref{table:setups}) as a function of time. Total magnetic field $|\vec{B}|$ is computed with an extended kernel of $1.5\Delta\text{x}_\text{cell}$ to avoid {\it X} and {\it O} points where $|\vec{B}| \sim 0$. Dashed lines correspond to maxima in each run at each given time, while solid lines represent averaged values. Red and grey lines display respectively the $|\vec{\nabla} \cdot \vec{B} / \vec{B}| \Delta\text{x}_\text{cell} = $1 and 0.1 ratios. Note that the maximal values in each simulation remain below a per cent, displaying the robustness of the CT scheme.}
\label{Divergences}%
\end{figure}

All our simulations are generated using the {\sc nut} initial conditions at
redshift $z = 500$ \citep{Powell11}. {\sc nut} is a cosmological cubic box of 9 $h^{-1}$ Mpc comoving on a side, with a 3 $h^{-1}$ Mpc comoving diameter spherical region carved out, where the zoom takes place. In the zoom sphere, a quasi-Lagrangian AMR strategy is allowed to refine the Eulerian grid down to 10 physical parsecs. The focus of our study is a $M_\text{vir} (z = 0) \simeq 5 \cdot 10^{11} M_\odot$ dark matter (DM) halo formed approximately at the centre of this zoom region, and the Milky Way-like galaxy it hosts. The mass resolution of the DM and stellar particles are $M_\text{DM} \simeq 5 \cdot 10^4 M_\odot$ and $M_{*} \simeq 5 \cdot 10^3 M_\odot$ respectively. Cosmological parameters are set accordingly to WMAP5 cosmology \citep{Dunkley09}. All our runs include an instantaneous UV background at $z = 10$ \citep{Haardt96}, and metal cooling above \citep{Sutherland93} and below \citep{Rosen95} a temperature of $10^4$ K. We initialise the simulation with a metallicity floor $Z = 10^{-3} Z_\odot$ to reproduce metal enrichment from the first stars \citep{Wise2012}. Gas is always assumed ideal, mono-atomic, and with specific heat ratio $\gamma = 5/3$. Magnetic fields are modelled employing supercomoving units \citep[following equation A26 in][]{Martel98}, and solved with a Constrained Transport (CT) method. CT ensures that the divergence of the magnetic field is kept to zero down to numerical precision. A common alternative approach to MHD is divergence cleaning. CT and divergence cleaning methods have been found to intrinsically lead to different results \citep{Balsara04}. Thus, simulation results concerning magnetic fields should be confirmed with various methods. One of the advantages of employing CT algorithms is that they ensure the absence of a non-physical monopolar component of the magnetic field, which can alter the evolution of the simulation \citep[see e.g.][where different magnetic solvers are compared for simple MHD test cases]{Hopkins16}. To illustrate this, in Fig. \ref{Divergences} we display the maximal (dashed) and average (solid) divergence to field ratio $|\vec{\nabla} \cdot \vec{B} / \vec{B}| \Delta\text{x}_\text{cell}$ for all cells in all simulations. Note that the dashed lines representing the maximal divergence relative to the magnetic field measured in each simulation remain below the percent level at all times. As other methods, CT algorithms introduce a numerical resistivity in the equations. For the {\sc RAMSES} implementation, this resistivity scales with the spatial resolution \citep{Teyssier06}. As the real magnetic diffusivity $\eta$ is negligible for galaxies and the intergalactic medium, we set its value to $\eta = 0$ in the induction equation (eq. (\ref{eq:Induction})), for all our simulations. All diffusive effects are therefore purely of numerical nature, arising from the indicated diffusivity inherent to the CT solver.

All our simulations employ a magneto-thermo-turbulent star formation model, already introduced in \citet{Kimm17, Trebitsch17, Mitchell18} and to be presented in more detail together with an analysis of its effects in Devriendt et al. (in prep). In this model, gas is converted into stars only within regions refined to the highest spatial resolution \citep{Rasera06} using a locally computed efficiency \citep{Padoan11, Federrath12} once gravity overcomes the combined support provided by the turbulent, magnetic and thermal pressures. We detail the extension we implemented of this star formation model to account for magnetic fields in Appendix \ref{ap:StarForm}. {In this work, the local magnetic field is not modified during star formation events.} While a subset of our simulations do not include stellar feedback (NoFb; NB20, NB12, NB11, and NB10), the rest of our runs have stellar feedback implementations that make use of a Kroupa initial mass function \citep[][IMF]{Kroupa01} to determine the ratio of supernovae (SN) per solar mass of stars formed. SN events take place 3 Myr after the formation of a stellar particle, injecting back to their host cell a gas mass fraction $\eta_\text{SN} = 0.213$, and a metal mass fraction $\eta_\text{metals} = 0.075$ when they occur. We only consider gas mass return and metal enrichment from SN. Furthermore, SN events do not inject magnetic energy, which will be addressed in future work (Martin-Alvarez et al. in prep). In those runs including stellar feedback, we implement one out of the two following models: either a mechanical supernova feedback (Mech) that better captures the expected momentum injection by SNe \citep[presented in][]{Kimm14, Kimm15}, or a purely thermal SN feedback with a simple representation of stellar radiation (RdTh) as used in \citet{Roskar14}. For the latter model we assume a large opacity of the gas in the ISM $\kappa_\text{IR} = 20 \text{ cm}^2 \text{ g}^{-1}$ \citep{Semenov03}, representing an ISM that efficiently absorbs dust infrared (IR) radiation. These two feedback prescriptions are described in further detail in \citet{Martin-Alvarez18}.

\subsection{Primordial magnetic fields}
\label{ss:InitialField}

The induction equation that governs the evolution of the magnetic flux density in MHD as a function of time, $\vec{B}\;(t)$, can be written as:
\begin{equation}
\frac{\partial \vec{B}}{\partial t} = \vec{\nabla} \times \left( \vec{v} \times \vec{B} \right)  +  \eta \vec{\nabla}^2 \vec{B},
\label{eq:Induction}
\end{equation}
where $v$ is the flow velocity and $\eta$ the magnetic diffusivity which vanishes for ideal MHD. In the absence of battery terms, this equation requires the presence of a magnetic seed to trigger any evolution of $B$. The theoretical uncertainties surrounding their origin translates into a large freedom to select the initial configuration of magnetic fields in cosmological simulations. Following common practice, we thus include initial magnetic seed fields which are uniform throughout the simulation volume. We explore a range of simple configurations by changing the amplitude of the primordial magnetic field and re-orienting it along one of the three axes of the simulated domain $( \vec{i}, \vec{j}, \vec{k})$. We note that different, more complex configurations can affect the evolution of simulations, particularly during galaxy collapse \citep{Zeldovich83, Rieder16}, and in the close environment of the galaxy. However, in this work, we restrict ourselves to the most elementary cases previously mentioned.

The employed configuration for the magnetic seed field is to be understood as the main component of a field coherent on $\sim$ comoving Mpc scales. This is consistent with some inflationary model predictions \citep{Ratra92, Barrow11, Sharma18}. As a word of caution we remind that these type of initial magnetic fields do not encapsulate any small scale fluctuations, which might have interesting local effects during the formation of galaxies \citep{Brandenburg05, Kandus11}. As perturbations collapse to form galaxies, the coherence of magnetic fields will break on scales of the order of the collapsing perturbation size. Indeed, even for a simplified setup such as the one we use, initial conditions such as the angular momentum of the region or turbulence generated on small scales alter the post-collapse magnetization of the medium \citep{Sur12}. In a cosmological scenario, compression, turbulence and numerical magnetic reconnection modify the energy spectrum, concentrating most of the magnetic energy on (sub)galactic scales. Whenever magnetic fields are dynamically important, they will also influence the collapse process itself. We plan to address how these different effects shape the circum-galactic medium at large in a follow-up publication. Finally, we remark that some of our results are due to the mere presence of magnetic fields in the galaxy, and could thus be extrapolated to other primordial, astrophysical, or battery modes of seeding.
\subsection{Simulations}
\label{ss:Simulations}
Our suite of simulations explores a variety of stellar feedback prescriptions, and magnetic field strengths and orientations. The entire set is summarised in Table \ref{table:setups}. The comoving primordial magnetic field strength $B_0$ is varied across the range $3 \cdot 10^{-20} \text{G} \leq B_0 \leq 3 \cdot 10^{-10}$ G. The B20 simulations ($B_0 \sim 10^{-20}$ G) feature an extremely weak field that does not affect the gas dynamics at any point of the run. From primordial magnetic field strengths of $B_0 \sim 10^{-14}$ G and above (B14 runs), our simulations reach magnetic fields in some regions of the proto-galaxy on the order of $\mu$G after the initial collapse. These are slowly amplified to permeate the rest of the galaxy. The B13 ($B_0 \sim 10^{-13}$ G), B12 ($B_0 \sim 10^{-12}$ G) and B11 ($B_0 \sim 10^{-11}$ G) runs display from collapse magnetizations in the entire galaxy on orders of $\sim 1 - 10\,\mu$G. Finally, the B10 runs are the most severe scenario we probe here, with a value of $B_0 \sim 10^{-10}$ G, still below the present observational upper limit provided by Planck, but close to it. In fact, $B_0 = 3 \cdot 10^{-10}$ G corresponds to a magnetic energy density $\sim 1$ dex below that predicted by the Planck+SPT upper limit for $B_0$. We explore three different orientations for each of the MB20, MB14, MB12, and MB10 fields. Due to the large computational cost of these simulations, all y-oriented ($\vec{j}$) runs are evolved to $z = 6$, x-oriented ($\vec{i}$) to $z = 4$, and only the z-oriented ($\vec{k}$) are evolved down to $z = 2$. Therefore, the accretion dominated phase \citep[$13 > z > 4$,][]{Martin-Alvarez18} can be studied with at least two runs for each primordial strength. It is during this phase that we expect the environment to have its largest impact on the evolution of the magnetism in the galaxy. To further examine how the importance of magnetic fields and stellar feedback compare, we also re-run some of these simulations changing the feedback prescription. All simulations with stellar feedback other than {\it Mech} have primordial magnetic fields aligned with the z axis ($\vec{k}$). Alternatively, all simulations with a non-z axis alignment employ the Mech feedback. Our fiducial model for stellar feedback, primordial magnetic field strength, and orientation has M (Mech), B12, and $\vec{k}$.

\begin{table}
\centering
\caption{Comoving primordial magnetic field strength and orientation ($\vec{B}_0$), 
stellar feedback model, and gas mass percentage with thermal to magnetic pressure ratio ($\beta$) lower than $10^3$ in the galactic region at $z = 6$ (defined in Section \ref{ss:HaloFinder}), for each run in the manuscript. We group simulations according to the strength of their initial magnetic field as B$X$ runs (with B$X$ indicating $B_0 (X) = 3 \cdot 10^{-X}$ G).}
\label{table:setups}

\begin{tabular}{l l l l}
\hline
\hline
Simulation & $\vec{B}_0$ (G)  & Feedback & $M_\text{gas}^\text{z = 6}\; (\beta < 10^3)$\\
\hline
\hline
{\textbf{B20 runs}} & & & \\
\textcolor{B20}{\textbf{MB20z}}  & $3 \cdot 10^{-20}\,\vec{k}$ & Mech & 0\% \\ 
MB20y & $3 \cdot 10^{-20}\,\vec{j}$ & Mech & 0\% \\
MB20x & $3 \cdot 10^{-20}\,\vec{i}$ & Mech & 0\% \\
NB20z & $3 \cdot 10^{-20}\,\vec{k}$ & \xmark & 0\% \\
RB20z & $3 \cdot 10^{-20}\,\vec{k}$ & RdTh & 0\% \\
\hline
{\textbf{B14 runs}} & & & \\
\textcolor{B14}{\textbf{MB14z}}  & $3 \cdot 10^{-14}\,\vec{k}$ & Mech & 4\% \\
MB14y & $3 \cdot 10^{-14}\,\vec{j}$ & Mech & 2\% \\
MB14x & $3 \cdot 10^{-14}\,\vec{i}$ & Mech & 6\% \\
\hline
{\textbf{B13 runs}} & & & \\
\textcolor{B13}{\textbf{MB13z}}  & $3 \cdot 10^{-13}\,\vec{k}$ & Mech & 26\% \\
\hline
{\textbf{B12 runs}} & & & \\
\textcolor{B12}{\textbf{MB12z}}  & $3 \cdot 10^{-12}\,\vec{k}$ & Mech & 63\% \\
MB12y & $3 \cdot 10^{-12}\,\vec{j}$ & Mech & 63\% \\
MB12x & $3 \cdot 10^{-12}\,\vec{i}$ & Mech & 72\% \\
NB12z & $3 \cdot 10^{-12}\,\vec{k}$ & \xmark & 80\% \\
RB12z & $3 \cdot 10^{-12}\,\vec{k}$ & RdTh & 64\% \\
\hline
{\textbf{B11 runs}} & & & \\
\textcolor{B11}{\textbf{MB11z}}  & $3 \cdot 10^{-11}\,\vec{k}$ & Mech & 95\% \\
NB11z & $3 \cdot 10^{-11}\,\vec{k}$ & \xmark & 99\% \\
RB11z & $3 \cdot 10^{-11}\,\vec{k}$ & RdTh & 97\% \\
\hline
{\textbf{B10 runs}} & & & \\
\textcolor{B10}{\textbf{MB10z}}  & $3 \cdot 10^{-10}\,\vec{k}$ & Mech & 99\% \\
MB10y & $3 \cdot 10^{-10}\,\vec{j}$ & Mech & 99\% \\
MB10x & $3 \cdot 10^{-10}\,\vec{i}$ & Mech & 100\% \\
NB10z & $3 \cdot 10^{-10}\,\vec{k}$ & \xmark & 100\% \\
RB10z & $3 \cdot 10^{-10}\,\vec{k}$ & RdTh & 99\% \\
\hline
\hline
\end{tabular}
\end{table}

\subsection{Measuring global properties in the galaxy}
\label{ss:GlobalProp}

\subsubsection{Finding dark matter haloes and galaxies}
\label{ss:HaloFinder}
To compute the position and properties of the dark matter haloes in our simulation, we apply the {\sc HaloMaker} software \citep{Tweed09} to the dark matter component. The galaxy is then identified by running the same algorithm on the baryonic component (gas and stars). Prior to the collapse, we fix the virial radius $r_\text{vir}$ to its physical (i.e. non-comoving) value when the halo is first found. We ensure the centre of the galaxy is accurately determined by computing it through recursive application of the shrinking spheres method \citep{Power03}. Once the centre has been identified, we compute properties of the identified structure such as ellipsoid of inertia axes and angular momentum (displayed in Fig. \ref{CollapseOrientation}).

We define the {\it galactic region} in each output using the physical virial radius $r_\text{vir} (t)$ of the dark matter halo. The galactic region is the spherical volume of radius $r_\text{gal} = 0.2\; r_\text{vir}$ centred on the position of the galaxy. It comprises the galaxy and its immediate surroundings, and its size thus increases with that of the galaxy and its dark matter halo.

\subsubsection{Global galactic properties}
\label{ss:Measurements}
Global galactic properties are measured within the entire galactic region ($r < 0.2\; r_\text{vir}$), unless otherwise indicated. We employ time-median measurements to smooth out temporary perturbations and concentrate instead on the secular evolution of the global properties. Measurements are made as follows: 
\begin{enumerate}
	\item At a given redshift of interest ($z_\text{target} =$ 10, 8, 6, 4, and 2), we estimate a dynamical timescale $\tau_\text{dyn}$, corresponding to the time required for a test particle to complete one full circular orbit with circular velocity
	\begin{equation}	
		v_\text{circ} (r) = \sqrt{\frac{G M (r)}{r}},
		\label{eq:vCirc}
	\end{equation}
	at radius $r = 0.2\; r_\text{vir}$,	where $M (r) = M_\text{g} (r) + M_\text{DM} (r) + M_* (r)$ represents the total mass contained within a spherical region of radius $r$, i.e. the sum of the gas ($M_g$), dark matter ($M_\text{DM}$) and stellar ($\Mst$) masses. The global properties of the galaxy should remain relatively unchanged for the duration of $\tau_\text{dyn}$, unless a disruptive event such as a merger takes place.
	\item For each simulation, we collect all available snapshots contained in the time interval of interest: $t (z_\text{output}) \in [t (z_\text{target}) - 0.5 \tau_\text{dyn}, t (z_\text{target}) + 0.5 \tau_\text{dyn}]$. Each data point contains a minimum of three snapshots.
	\item The value of each quantity (and associated errors) measured over the target time interval is chosen to be the time-weighted median (inter-quartile range). Each snapshot has a weight equal to the fraction of the target time interval represented by the snapshot. The time of transition between snapshots is taken as the equidistant time between the two outputs.
\end{enumerate}

Note that by computing global quantities in this manner, the quoted error bars do not strictly reflect the uncertainty of the estimate, but also represent a measure of its time variation and non-secular changes during $\tau_\text{dyn}$. Consequently, error bars yield information about the variability of the measurement over this timescale. We examined projected maps of the gas and stellar densities for various data points, and found that those possessing large error bars are mostly associated with non-secular events such as mergers.

\subsection{Mock imaging with Sunset}
\label{ss:Sunset}
We assess the observational impact of magnetic fields on the appearance of galaxies by looking at synthetic face-on projections of them. These are generated with the {\sc sunset} code, a simplified version of the {\sc stardust} algorithm presented in \citet{Devriendt99}. For all stellar particles in cubic boxes of size $\left(8~\text{kpc} \right)^3$ centred on the galaxy, we compute their emission in the corresponding James Webb Space Telescope (JWST) NIRcam filters\footnote{\href{https://jwst-docs.stsci.edu/display/JTI/NIRCam+Filters}{https://jwst-docs.stsci.edu/display/JTI/NIRCam+Filters}} $\left[\text{F090W}, \text{F150W}, \text{F356W}\right]$. These filters correspond to $UVJ$ rest-frame for a galaxy at $z=2$. Each stellar particle is treated as a single stellar population, modelled according to \citet{Bruzual03}. We follow \citet{Kaviraj17} and model dust as an absorption column, with a metal to dust mass ratio of 0.4. This value is found to be a good approximation to a more elaborate radiative transfer treatment \citep[see][]{Kaviraj17} .

To perform colour and concentration measurements of galaxies, we compute the Petrosian radii $\mathcal{R}_P$ \citep{Blanton01} and Petrosian fluxes within $2\mathcal{R}_P$ for each image. Concentration parameters are defined as $C_{50} = r_{50} / r_{90}$ where $r_X$ is the radius within which $X$\% of the Petrosian flux is contained.

\section{Results}
\label{s:Results}
Within the range of primordial magnetic fields allowed by current constraints, two main regimes can be distinguished from a galaxy formation perspective: weak primordial magnetic fields which do not lead to $\sim \mu$G magnetisation during the collapse of the proto-galaxy, and strong primordial magnetic fields for which this magnetisation level is reached shortly after collapse.

In the low primordial magnetic field scenario (with comoving strength $B_0 \lesssim 10^{-14}$ G), galactic dynamo processes are expected to amplify the galactic magnetic field to $\mu$G levels \citep{Pakmor14,Martin-Alvarez18}. As a consequence, no information regarding the initial state of the magnetic field can be recovered from the galaxy and the specific value of the primordial strength becomes relatively unimportant. However, this is not the case for primordial magnetic fields with much larger strengths ($B_0 > 10^{-14}$ G). These magnetic fields provide the expected $\approx \mu$G magnetisation of the galaxy through the compression of magnetic field lines during the proto-galaxy collapse phase. Throughout the literature, it is consistently pointed out that such strong fields would have a non-negligible impact on the general population of galaxies and structure formation \citep{Tsagas00, Marinacci15, Varalakshmi17, Safarzadeh19}.

The first process in the formation of our simulated galaxy is the collapse of the original density perturbation. Collapse, folding, and compressive motions alter the magnetic properties of the galaxy. Depending on the geometry, different arrangements between magnetic field lines and velocity flows will develop \citep{Zeldovich83}. During later cold gas accretion, the main supply of gas to a galaxy is anisotropic. As such, one expects the geometry of cosmological structures to interact with the intrinsically vectorial cosmic magnetic fields. The effect of local relative orientation should also be more significant when the coherence scale of the cosmic magnetic field is comparable to the length scale of density perturbations like filaments and walls, particularly when employing non-uniform magnetic initial conditions. 

\subsection{The influence of primordial magnetic fields on the initial collapse of the galaxy}
\begin{figure*}%
\includegraphics[width=2\columnwidth]{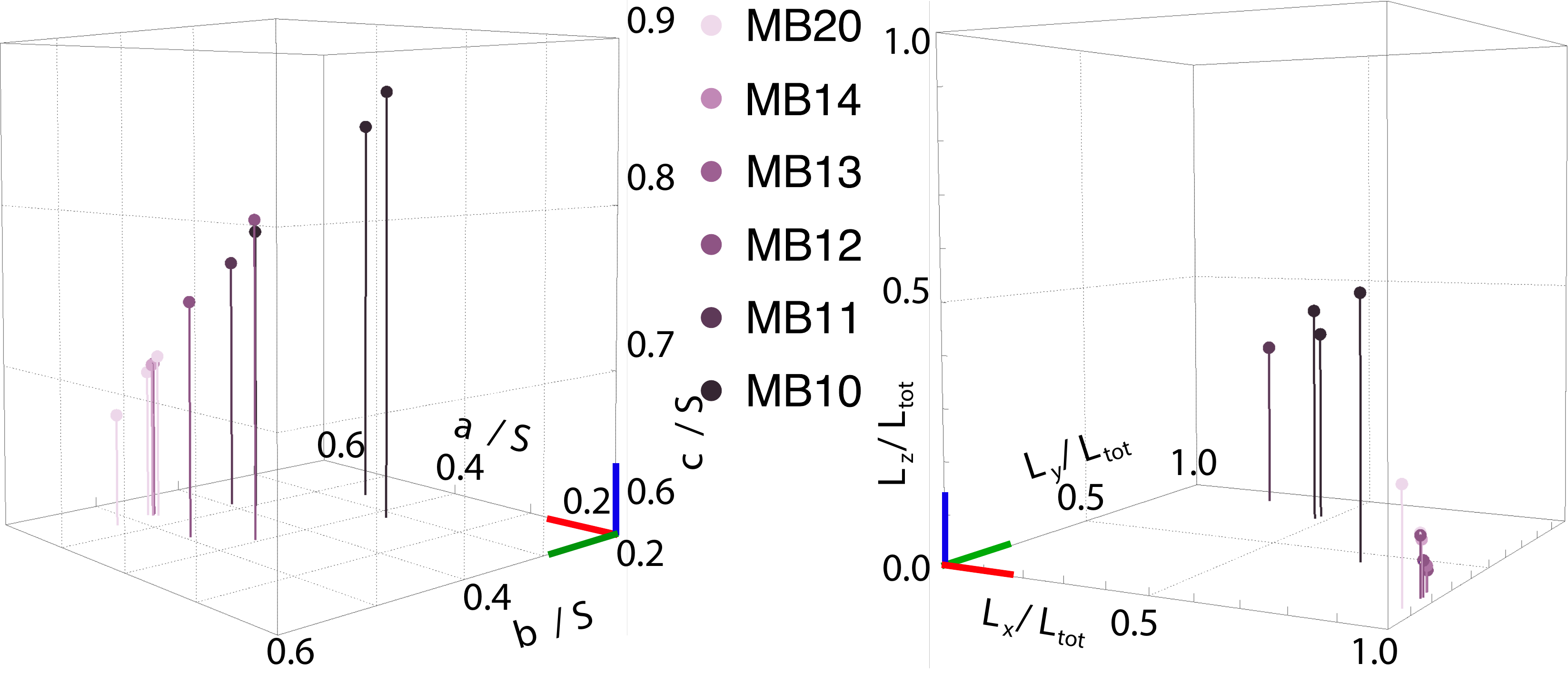}
\caption{Geometric properties of the galaxy at collapse ($t_\text{coll}$) for all primordial magnetic field orientations and $B_0$ strengths (represented by the colours indicated in the legend) used in the MB$X$ runs. (Left) Normalized baryonic ellipsoid of inertia axes ($S = \sqrt{a^2 +  b^2 + c^2}$). (Right) Normalized angular momenta of the baryonic component. As magnetisation increases, the shape of the galaxy elongates. The RGB coloured trihedra indicate the direction of increase along each of the corresponding abc-axes (left) and xyz-axes of the box (right). For the strongest $B_0$ probed (MB12, MB11 and MB10 runs), magnetic fields have an impact on the process of collapse.}
\label{CollapseOrientation}%
\end{figure*}

To better illustrate the implications of the primordial magnetic field strength, we show in Fig. \ref{CollapseOrientation} the normalized triaxial dimensions of the baryonic ellipsoid of inertia of the galaxy (left panel), and its angular momentum (right panel) for a series of simulations with Mech feedback and varying primordial magnetic field strength and orientation. These are measured by applying {\sc HaloMaker} to the baryonic component at the end of the collapse phase ($t_\text{coll}$). This moment is defined as the time when the galaxy has formed a significant amount of stellar particles (i.e. a minimum of 100). As shown in Fig. \ref{CollapseOrientation}, the galaxy collapses in all runs to similar shapes by $t_\text{coll} \sim 0.34 - 0.36$ Gyr. Only for the most extreme primordial magnetic field (MB10 simulations) is the collapse slightly delayed to $t_\text{coll} \sim 0.4$ Gyr. The semi major axis $c$ becomes larger with respect to the others as the strength of the primordial magnetic field increases. The other two axes have similar sizes ($a \sim b$) and the galaxy morphology resembles a prolate spheroid. From $B_0 \gtrsim 10^{-12}$ G, evidence of the magnetic field influencing the shape becomes more pronounced.

Angular momenta display a similar behaviour to the shapes. For most primordial strengths probed the angular momentum vector of the galaxy at $t_\text{coll}$ is contained in the plane defined by the xy axes of the box. Normalized galaxy angular momenta yield $\vec{L} / \| \vec{L} \| \sim \left(0.97, 0.22, 0.11\right)$ in most $\vec{B}_0$ orientations for strengths MB20, MB14, MB13, and MB12. For MB11 and MB10, the relative contribution of $L_x$ is decreased and more significant values are found for the other two components. Any temporary modifications of the angular momentum induced by magnetic fields are not expected to modify the post-collapse morphology \citep{Sur12} significantly as after the proto-collapse has ended, the angular momentum evolution of the galaxy becomes rapidly dominated by cold accretion flows \citep{Kimm11,Tillson15}.

The impact of the primordial magnetic field is the most evident as their strength, $B_0$, is increased. As our {\it MB20} simulations have a negligible magnetic energy, below the turbulent and thermal energies (see Table \ref{table:EnergyRatios}) by at least 15 orders of magnitude, we use them as a reference for the no magnetic field case. The effect of magnetic fields remains marginal for $B_0 = 10^{-14}$ G ({\it MB14} runs), with collapse only leading to $\sim \mu$G magnetic fields in part of the galaxy. As a result, only small deviations with respect to the MB20 runs are measured. However, as the magnetic field amplitude is increased beyond $10^{-14}$ G, the thermal to magnetic pressure ratio, $\beta$, decreases for a larger fraction of the galaxy and amplification through simple compression is reduced. The {\it MB13} run starts to display a small amount of variation in galaxy angular momentum, and {\it MB12}, {\it MB11}, and {\it MB10} runs have magnetic fields strong enough to alter the shape of the proto-galaxy and in the case of {\it MB10}
delay the collapse by tens of Myr.   

Once the initial perturbation has collapsed ($z \sim 13$), the growth of the galaxy will be mainly governed by cold gas supplied by cosmic filaments. We refer to this period as the accretion phase ($13 > z > 4$), and study the galaxy shortly after its collapse at the very early stages of this phase ($z = 10$). A higher strength of the primordial magnetic field implies that the magnetic energy supplied during the accretion of magnetised pristine gas could be non-negligible in comparison with that in the ISM of the galaxy, especially given that dynamically important magnetic fields can back-react on dynamo amplification and reduce its growth rate. In this regime, we suspect that not only the strength of the primordial magnetic field will affect the properties of galaxies, but that there are also effects related to its local orientation and its spatial coherence length.

\begin{table}
\centering
\caption{Gas energies comparison in the galactic region at $z = 6$ for different strengths of the primordial magnetic field. The Table only gives values for the {\it Mech} feedback runs with $\hat{k}$ oriented $\vec{B}_0$. Columns indicate for each run the specific magnetic energy $\varepsilon_\text{mag}$ to specific thermal energy 
$\varepsilon_\text{th}$ ratio, the specific turbulent energy $\varepsilon_\text{mag}$ to specific thermal energy ratio, and the specific thermal energy respectively.}
\label{table:EnergyRatios}
\begin{tabular}{l l l l}
\hline
\hline
Simulation & $\varepsilon_\text{mag}/\varepsilon_\text{th}$ &  $\varepsilon_\text{turb}/\varepsilon_\text{th}$ &  $\varepsilon_\text{th} (\text{erg}/\text{g})$\\
\hline
\hline
\textcolor{B20}{\textbf{MB20z}} & $1.5 \cdot 10^{-17}$   & $22.2$ & $1.4 \cdot 10^{12}$ \\
\textcolor{B14}{\textbf{MB14z}} & $2.9 \cdot 10^{-5}$    & $67.3$ & $4.4 \cdot 10^{11}$ \\
\textcolor{B13}{\textbf{MB13z}} & $0.2 \cdot 10^{-3}$    & $33.9$ & $8.2 \cdot 10^{11}$ \\
\textcolor{B12}{\textbf{MB12z}} & $0.05$                 & $16.7$ & $1.6 \cdot 10^{12}$ \\
\textcolor{B11}{\textbf{MB11z}} & $1.03$                 & $23.3$ & $1.2 \cdot 10^{12}$ \\
\textcolor{B10}{\textbf{MB10z}} & $3.62$                 & $13.0$ & $4.4 \cdot 10^{12}$ \\
\hline
\hline
\end{tabular}
\end{table}

In light of this discussion, as well as the measured mass fraction of significantly magnetised gas (Table \ref{table:setups}), and the measured magnetic energy ratios\footnote{For each form of energy $X$, the specific energy $\varepsilon_X$ is the ratio of the absolute energy $E_X$ to gas mass. Each $E_X$ is calculated by computing the contribution to the absolute energy on a cell-per-cell basis. For each cell, the turbulent energy is computed by removing from the tangential component of the velocity vector the circular velocity \citep[see][for additional detail on how we perform this calculation]{Martin-Alvarez18}.} (Table \ref{table:EnergyRatios}) in the galactic region, we expect each group of runs to behave as follows: B20 runs will display virtually no magnetic effect; B14 runs will only show small deviations from B20; the B13 run will constitute an intermediate case where the effect of magnetic fields start to manifest more notably. Finally, the B12, B11 and B10 simulations, which already display a clear impact of the primordial magnetic field by the end of the collapse phase, will exhibit the largest departure from the B20 runs. We quantify these differences in Section \ref{ss:GalacticPhysics}, devoted to global morphological and dynamical properties of galaxies, and explore the role played by magnetic braking in driving them. Section \ref{ss:LowZ}, discusses how the presence of strong primordial magnetic fields at high redshift would manifest itself in simple $UVJ$ observations of galaxies.

\subsection{Impact on global galaxy properties}
\label{ss:GalacticPhysics}

\subsubsection{Morphology}
\label{sss:Morphology}
\begin{figure*}%
    \includegraphics[width=2\columnwidth]{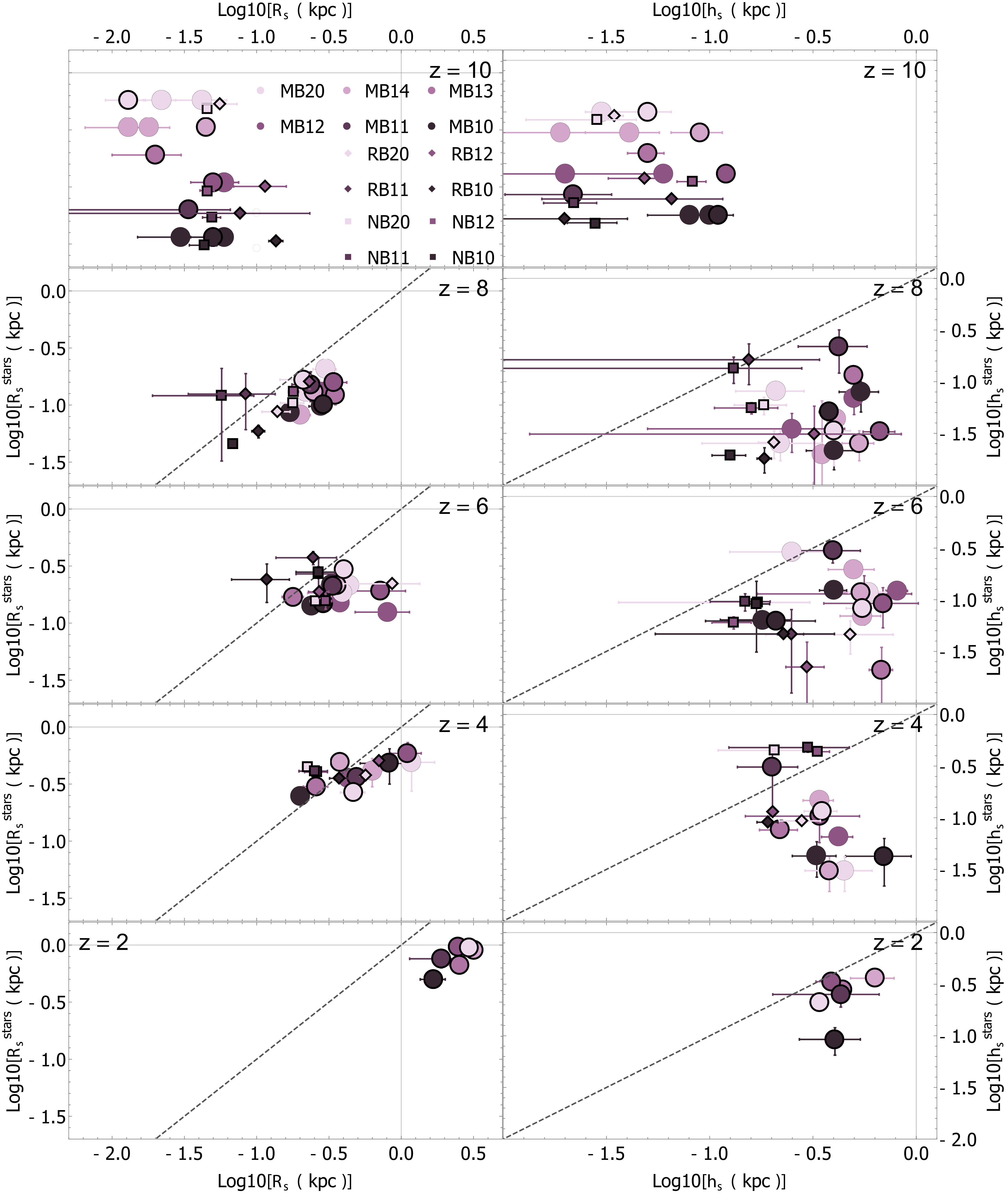}\\
    \caption{Changes in the galaxy stellar radial scale length $R_s^\text{stars}$ vs. gas radial scale length $R_s$ (left column) and stellar scale height $h_s^\text{stars}$ vs. gas scale height $h_s$ (right column) as a function of primordial magnetic field strength. The field strength is represented by the colour of the data points, with $B_0$ increasing as the shade of purple darkens. Circles ({\it Mech}), diamonds ({\it RdTh}), and squares ({\it NoFb}) correspond to different feedback prescriptions. Data points associated to {\it Mech} runs with $\vec{B_0}$ oriented along the box $\vec{k}$ direction are highlighted by a thicker symbol contour line. Finally, panels from top to bottom display decreasing redshifts: $z =$ 10, 8, 6, 4, and 2. Note that for $z = 10$, neither $R_s^\text{stars}$ nor $h_s^\text{stars}$ are shown due to the absence of a significant galactic stellar component. Error bars represent the interquartile range and are computed as indicated in Section \ref{ss:Measurements}. The dashed line corresponds to $R_s^\text{stars} = R_s$ for the left panels, and $h_s^\text{stars} = h_s$ for the right panels. We find the gas disk to be larger in both radial scale length and vertical scale height than the stellar disk. At $z = 2$, both radial scales are reduced as $B_0$ increases.}%
    \label{GalScales}%
\end{figure*}

The global morphology of a disk galaxy can be defined by two numbers: its vertical scale height $h_s$ and radial scale length $R_s$. In order to understand whether primordial magnetic fields have an impact on the morphology of our simulated galaxies, we thus review how these two parameters change over redshift as a function of primordial field strength and configuration. The $h_s$ and $R_s$ scales are computed by fitting exponential density profiles separately for both the gaseous and stellar components. Appendix \ref{ap:MorphFit} details this fitting procedure and showcases some generic examples. We display the variation of the gas-vs-stellar radial scale length (left panel) and vertical scale height (right panel) in Fig. \ref{GalScales}, as a function of increasing primordial magnetic field strength. Each row of panels corresponds to subsequently decreasing redshifts, as indicated on each panel. Different symbols represent runs with different feedback prescriptions, and the strength of the primordial magnetic field in each run is indicated by the colour of the points, becoming darker as the strength of the magnetic field is increased. Data points corresponding to runs with a $\hat{k}$-oriented primordial magnetic field are outlined with a thicker line and are the only ones available all the way down to $z = 2$. Dashed lines correspond to equal scale lengths for the stellar and gaseous components.

While it is natural to expect magnetic fields to induce changes in the morphology of the gaseous component, given the absence of a direct interaction between the stellar component and the magnetic field, one might expect the stellar component to be by-and-large unaffected. However, it is still possible for the magnetic field to influence the stellar component indirectly by intervening in the process of star formation, especially at high redshift, when the stellar population is relatively young.

The two topmost panels of Fig. \ref{GalScales} show the impact of different primordial magnetic fields on the galaxy morphology shortly after collapse ($z = 10$). At this stage, given our stellar particle mass resolution ($M_* \sim 5 \cdot 10^3 M_\odot$) the stellar component is not resolved enough to accurately determine the galaxy morphology, and thus it is not displayed. Similarly, due to the absence of a well-defined rotationally-supported disk and an irregular shape of the galaxy at this redshift, $h_s$ is a measure of the gas scale length along its main rotation axis. From looking at the figure, one can clearly see the trend that an increase of the primordial magnetic field strength leads to increased support against collapse, thus yielding larger gaseous scale lengths. Both $R_s$ and $h_s$ are nearly doubled when comparing the strongest $B_0$ runs, {\it MB10} with {\it MB20}. In the absence of feedback (square symbols), the effects of $B_0$ are less obvious. The {\it RdTh} feedback model runs (diamond symbols) lead to the largest radial scale length of the galaxy, but at the expense of a reduction in thickness for the strongest magnetic field.

During the rest of the accretion phase ($10 > z > 4$), opposing trends to those obtained immediately after collapse emerge. $R_s$ shifts from increasing with $B_0$ to decreasing. A similar behaviour is observed for the stellar component $R_s^\text{stars}$. This behaviour starts to manifest at $z = 8$ and is strongly asserted by $z = 2$, well after the end of the accretion phase and once the galaxy acquires a fully developed rotationally-supported gas disk. Note that this trend is not immediately apparent, as some runs display large deviations from the bulk of the distribution of data points with the same $B_0$. These also display significantly larger logarithmic error bars, which are associated in the majority of the cases with merger events temporarily disturbing the morphological appearance of the galaxies. Examples are {\it RB20z}, {\it RB10z}, {\it MB12z}, and {\it MB12y} at $z = 6$, or {\it MB20x}, {\it MB12z} and {\it MB10z} at $z = 4$. At $z = 2$, the radial scale length of the gas disk is halved from 3 kpc in the absence of significant magnetic field ({\it MB20z} run), to 1.5 kpc for the highest $B_0$ ({\it MB10z} run). Interestingly, a larger relative reduction of this scale is found for the stellar component, which we attribute partially to a more centrally compact distribution of the star formation as $B_0$ is increased (see Section \ref{ss:LowZ}, Fig. \ref{SFRscales}). Once again, in the absence of feedback, $B_0$ seems to not alter the radial scale length significantly between {\it NB20z}, {\it NB12z}, and {\it NB11z} at any redshift $z \gtrsim 4$. 

The effect of $B_0$ on the vertical scale height is hard to establish before $z \sim 10$ due to the limited spatial resolution of our numerical simulations. During the accretion phase, significant turbulence is driven by accretion-related processes \citep{Elmegreen10, Klessen10}. These processes provide enough support to establish the gas disk scale height at approximately $h_s \sim 300 - 400$ pc, albeit with a large spread, especially at high redshift ($z > 4$). This leads to an apparent lack of sensitivity of $h_s$ to $B_0$. If anything, intermediate primordial magnetic fields ({\it B14}, {\it B13}, {\it B12}, {\it B11}) seem to produce thicker $h_s$. By $z = 2$, the presence of dynamically important magnetic fields leads to an extra thickening of the rotationally-supported gaseous disk of at most 100 pc as compared to {\it MB20z}. As redshift $z = 2$ is reached, all runs but {\it MB10z} seem to converge towards equality between $h_s$ and $h_s^\text{stars}$.  It is possible the observed reduction of $h_s^\text{stars}$ in {\it MB10z} has some connection to the inclusion of magnetic pressure in our star formation algorithm. In this prescription, star forming clumps have to reach higher masses to overcome the additional magnetic pressure (see appendix \ref{ap:StarForm}). This gas accumulation is more likely to take place close to the disk mid-plane, where the gas density is higher. Finally, there exists a more significant scatter across the vertical scale height panels than across the radial scale length ones. A more detailed analysis of the influence of magnetic fields on the disk thickness seems to indicate that this is likely caused by disk flaring, but this is beyond the scope of this work, aimed primarily at measuring global properties.

While the magnetic fields at play in these simulations are of primordial origin, changes in the vertical scale height arise solely from magnetic pressure, and therefore they are expected to also occur when a magnetisation of similar amplitude is produced by other mechanisms (e.g. feedback from compact sources). Whether this is also the case for the radial scale length is unclear as the impact on this quantity is related to how magnetic fields affect the angular momentum of the galaxy. We address this issue in the next section.  

\subsubsection{Dynamics}
The presence of magnetic fields in the ISM has been reported to affect its dynamics: from turbulence modes and scales \citep{Kinney00, Schekochihin04, Zamora-Aviles18}, to the amount of gas collapsing into molecular clouds and star forming regions \citep{Hennebelle14,Hull17}, or the gas mass fraction present in the various ISM phases \citep{Villagran17}. On larger scales, magnetic fields could influence global dynamical properties such as the gas rotational velocity of galaxies. Indeed, through magnetic braking and angular momentum transport, they could reduce galactic rotation and establish inward gas flows \citep{Sparke82, Beck15}. Alternatively, \citet{Ruiz-Granados10} suggest that magnetic fields could boost galactic circular velocities at large radii.

In this section, we address  how magnetic fields influence the global dynamical properties of our simulated galaxy. We quantify small-scale dynamical changes by focusing on a single number: the total turbulent velocity dispersion $\sigma_\text{rms}$. It is computed in a spherical coordinate system ($r, \theta, \phi$), co-moving with the galaxy, as the root of the sum of the squared mass-weighted average deviations of each component of the velocity from the mass-weighted average velocity in thin shells of radius $r$.

Similarly, to study changes in the dynamics on galactic scales, we quantify the global rotation of the galaxy using the spin parameter. This parameter indicates the degree of rotational support of a given component (stars or gas). For each of these components $i$, it is defined as
\begin{equation}
    \lambda^i_\text{rot} = \frac{L_i}{\sqrt{2} r M (r) v_\text{circ} (r)} \simeq \frac{L_i E_B^{1/2}}{G M (r)^{5/2}},
    \label{eq:RotPar}
\end{equation}
following \citet{Bullock01}. In equation (\ref{eq:RotPar}) the radius used is that of the entire galactic region $r = 0.2 r_\text{vir}$, $L_i$ is the total angular momentum of the $i$ component in the region, $E_B$ is the binding energy, $G$ is the gravitational constant, and $M (r)$ is the total total mass enclosed in the region. 

\begin{figure*}%
    \includegraphics[width=2\columnwidth]{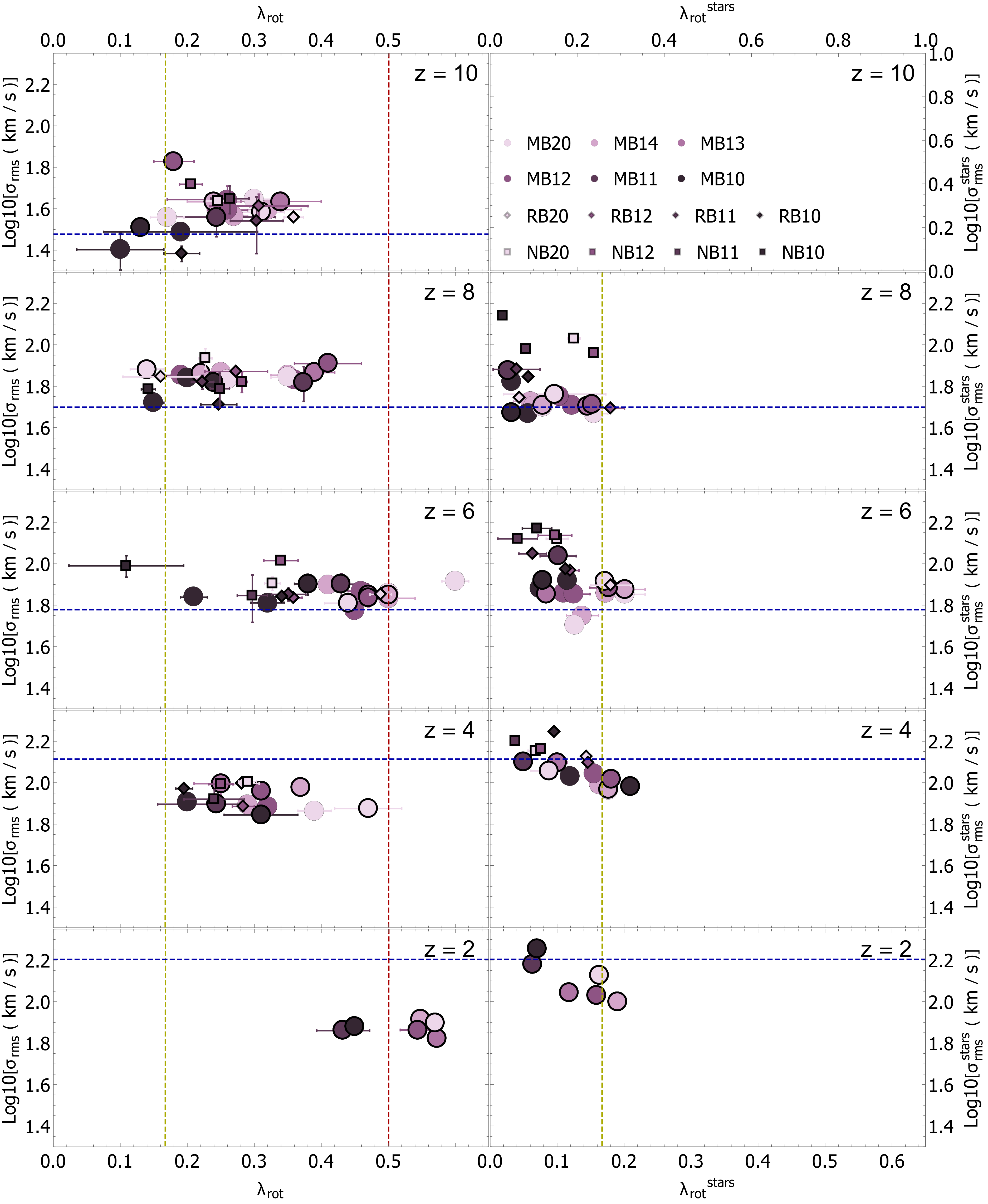}\\
    \caption{Changes of the gas dynamical ($\sigma_\text{rms}$ vs $\lambda_\text{rot}$; left column) and stellar dynamical ($\sigma_\text{rms}^\text{stars}$ vs $\lambda_\text{rot}^\text{stars}$; right column) properties of the galaxy when increasing the primordial magnetic field $B_0$. Data points legend (shown at the top right of the figure) is as for Fig. \ref{GalScales}. Vertical lines correspond to an approximate rotational support threshold $\lambda_\text{rot} = 0.5$ (red) and one third of this value (olive). The horizontal blue line represents instead the average circular velocity of the {\it MB} runs at 0.2 $r_\text{vir}$. Stellar component quantities are not shown for $z = 10$ due to the lack of enough stellar particles in the galaxy. Gas becomes more rotationally supported as redshift decreases, and the ratio of gas velocity dispersion to circular velocity falls. In contrast, stars do not show large support by coherent rotation. Gas velocity dispersion has small scatter with magnetic field strength and feedback prescription. However, there is a trend for runs with stronger $B_0$ to have both lower gas and stars spin parameters.}
    \label{GalDyns}%
\end{figure*}

As done for the radial scale lengths and vertical scale heights, we plot in Fig. \ref{GalDyns} the evolution of these quantities for the gas and stellar components, starting shortly after collapse, going through the entirety of the accretion phase and further in the feedback dominated phase down to $z = 2$. We separate the dynamics of each baryonic component by presenting $\sigma_\text{rms}$ vs $\lambda_\text{rot}$ (left column) and $\sigma^\text{stars}_\text{rms}$ vs $\lambda^\text{stars}_\text{rot}$ (right column) for gas and stars respectively. 

For the gaseous component, stronger primordial fields cause a clear and significant reduction of $\lambda_\text{rot}$, especially at $z < 8$. This reaches up to an absolute decrease of $\Delta \lambda \sim 0.2$ and appears quite independent of the stellar feedback model or the absence of feedback altogether. At $z = 2$, the spin parameter for the runs with $B_0 \leq 10^{-12}$~G stabilises well above the minimal threshold for rotational support ($\lambda_\text{rot} \gtrsim 0.5$, red dashed vertical line), and is only slightly changed by $B_0$. In contrast, {\it MB11z} and {\it MB10z} exhibit a significantly lower contribution from rotation to support against gravity.

\begin{figure}%
	\includegraphics[width=\columnwidth]{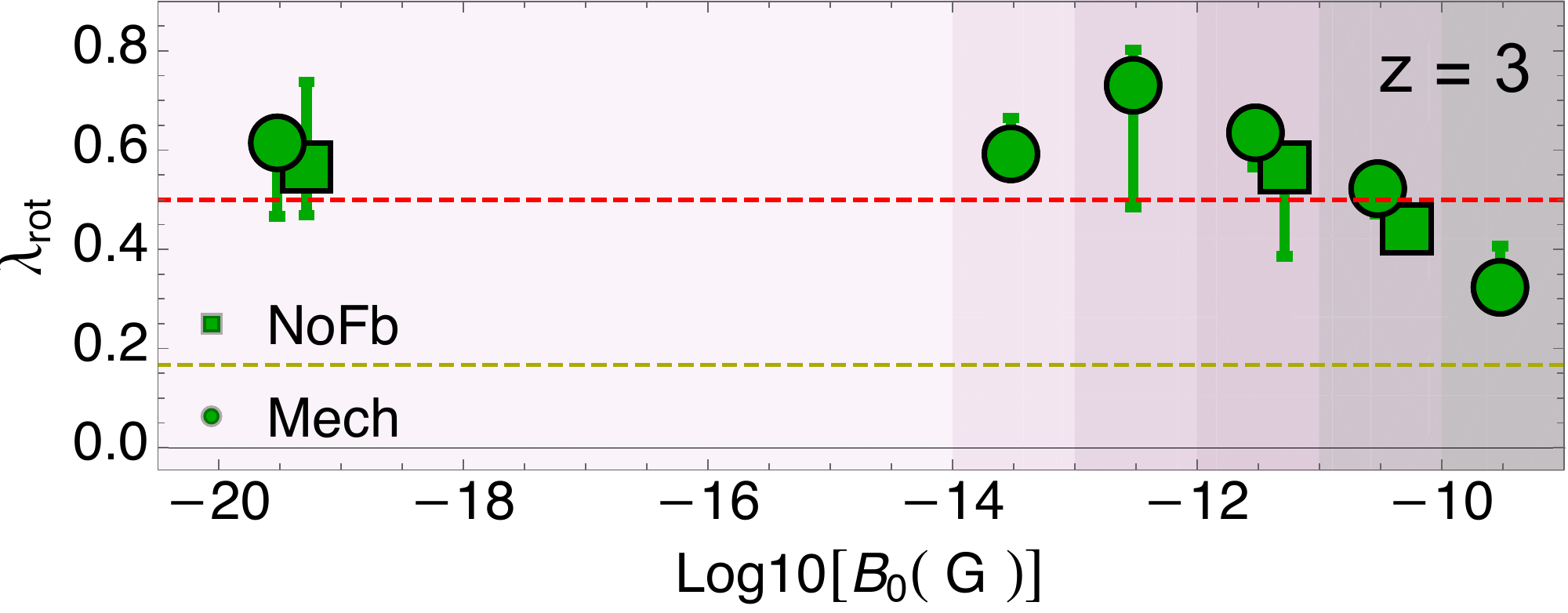}\\
	\caption{Changes of the gas spin parameter $\lambda_\text{rot}$ of the galaxy vs $B_0$ at $z = 3$ (i.e. the lowest redshift reached by {\it NB} simulations). {\it NoFb} data points have been slightly displaced in the x-axis to improve readability. Error bars as for Fig. \ref{GalScales}. Dashed lines correspond to an approximate rotational support threshold $\lambda_\text{rot} = 0.5$ (red) and one third of this value (olive). Runs with $B_0 > 10^{-11}$ G display lower $\lambda_\text{rot}$ values.}%
	\label{Z3rotation}%
\end{figure}

Efficient stellar feedback is considered to assist in establishing rotational support for the gas in galaxies, facilitating the formation of extended disks \citep{Scannapieco2008,Ceverino2017}. As a result, an interesting question is how do stellar feedback and $B_0$ interact when contributing to the final $\lambda_\text{rot}$ of a galaxy. To address this, we further evolve the no feedback simulations {\it NB20}, {\it NB12}, and {\it NB11} down to $z =3$ and display in Fig. \ref{Z3rotation} their $\lambda_\text{rot}$ in combination with the $\lambda_\text{rot}$ measurements from the {\it MBz} runs. In agreement with a scenario where stellar feedback contributes to rotational support, we find runs with no feedback display lower $\lambda_\text{rot}$ at a fixed $B_0$. We would expect the divergence between $\lambda_\text{rot}$ in the two sets of simulations to increase as galaxies continue evolving towards lower redshift \citep{Ceverino2017}. Similarly, we find evidence for lower $\lambda_\text{rot}$ values to occur in the galaxies with $B_0 > 10^{-11}$ G. The trend also seems to be preserved in the absence of feedback, although we note the comparison is only made for three {\it NB} runs.

During the accretion phase ($z \gtrsim 4$), there is no evidence that changing the stellar feedback prescription or primordial magnetic field has a significant impact on the gas velocity dispersion $\sigma_\text{rms}$. This is somewhat expected: turbulence during this phase is dominated by accretion related processes, be it through direct energy injection \citep{Klessen10}, or gravitational instabilities \citep{Elmegreen10}. Similarly, stellar feedback is regulated by gas infall \citep{Hopkins13}. We also find that most of the merger events that alter the morphological properties do not frequently have a large impact on the dynamics, neither by yielding values of $\sigma_\text{rms}$ or $\lambda_\text{rot}$ completely at odds with their usual non-merging distribution, nor by significantly stretching their dispersion. At $z = 4$, there might be some minor evidence suggesting that for a given $\lambda_\text{rot}$, stronger primordial magnetic fields lead to a small reduction of $\sigma_\text{rms}$, but it remains marginal at best.

As the support of the stellar component is purely dynamic, and once heated this collisionless component cannot cool, it displays a lower level of rotational support than the gas and hence lower spin parameters $\lambda_\text{rot}^\text{stars}$. Note that values of $\lambda_\text{rot}^\text{stars}$ would be much higher (comparable to $\lambda_\text{rot}$) if one considered only the young (star particles with ages $\lesssim 100$ Myr) stellar population. The value of $\lambda_\text{rot}^\text{stars}$ remains close to the vertical olive dashed line in Fig. \ref{GalDyns}, which indicates a value corresponding to a third of the minimal threshold for rotational support. The stellar spin parameter also manifests, albeit to a lesser degree than for the case of the gaseous component, some reduction as $B_0$ increases. This follows from the observed reduction of this gaseous spin parameter, to which it couples through the process of star formation. The trend is notably clear for the {\it NoFb} runs. Stellar spin parameters are affected to some degree by the feedback prescription selected, but changes appear somewhat stochastic. 

Interestingly unlike $\sigma_\text{rms}$, $\sigma_\text{rms}^\text{stars}$ displays a clear trend as magnetisation increases. This is because $\sigma_\text{rms}^\text{stars}$ is more directly linked to $\lambda_\text{rot}^\text{stars}$ as the collisionless stars cannot cool: if the coherent rotation of the stellar component is reduced, its turbulent component has to increase correspondingly to maintain dynamical support against gravity. This anti-correlation can best be observed in Fig. \ref{GalDyns} at $z \lesssim 4$. $\sigma_\text{rms}^\text{stars}$ is also affected by the stellar feedback prescription (or its absence). Both {\it RdTh} and {\it NoFb} runs display higher $\sigma_\text{rms}^\text{stars}$ than {\it Mech} runs, accumulated on top of the increase caused by $B_0$. This could be the result of a less efficient stellar feedback producing a larger stellar spheroidal component \citep{Scannapieco2008}, reflected by higher $\sigma_\text{rms}^\text{stars}$ values.

Magnetic fields are expected to affect gas turbulence after the accretion phase, even in the case when they are not of primordial origin, because this latter is sensitive to magnetic pressure on small scales. On the other hand, a decrease in the spin parameter will depend on whether the mechanism causing the loss of angular momentum still operates when magnetic fields are generated on smaller scales. We now proceed to review this process in more detail.

\begin{figure*}%
	\includegraphics[width=1.5\columnwidth]{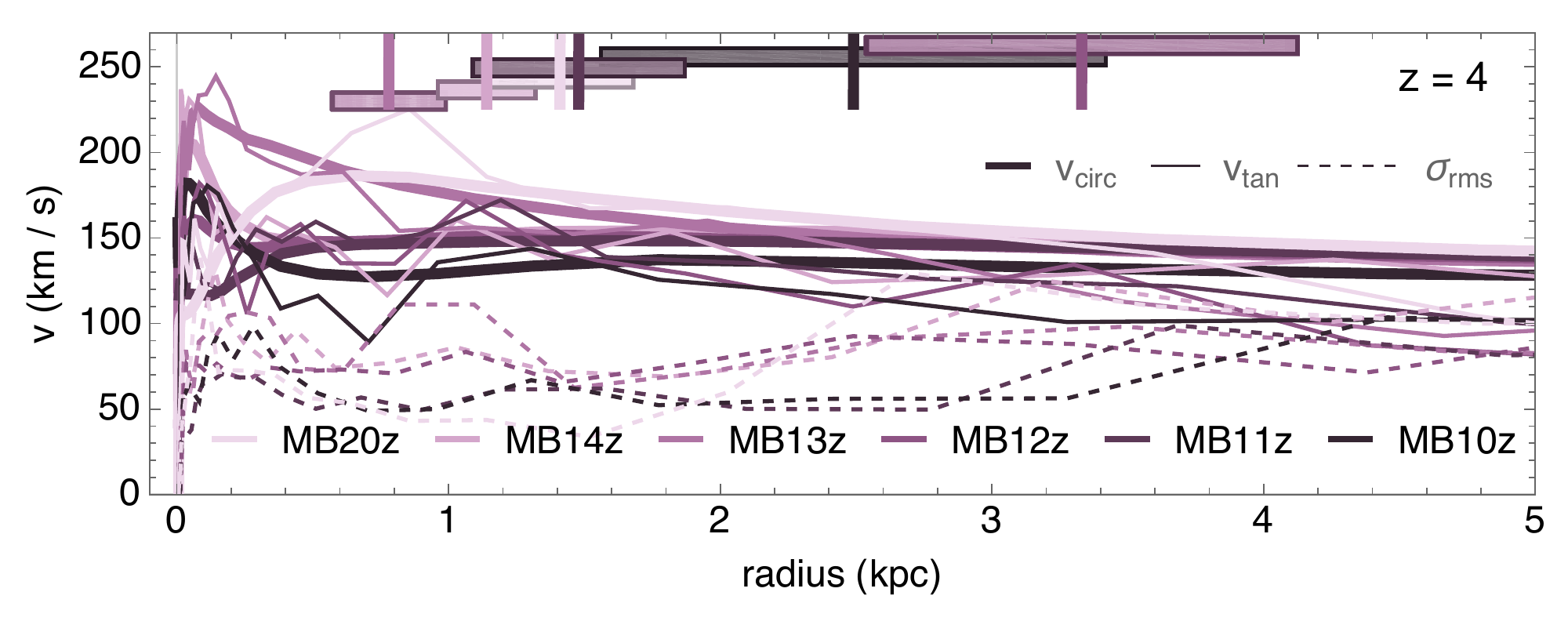}\\
	\includegraphics[width=1.5\columnwidth]{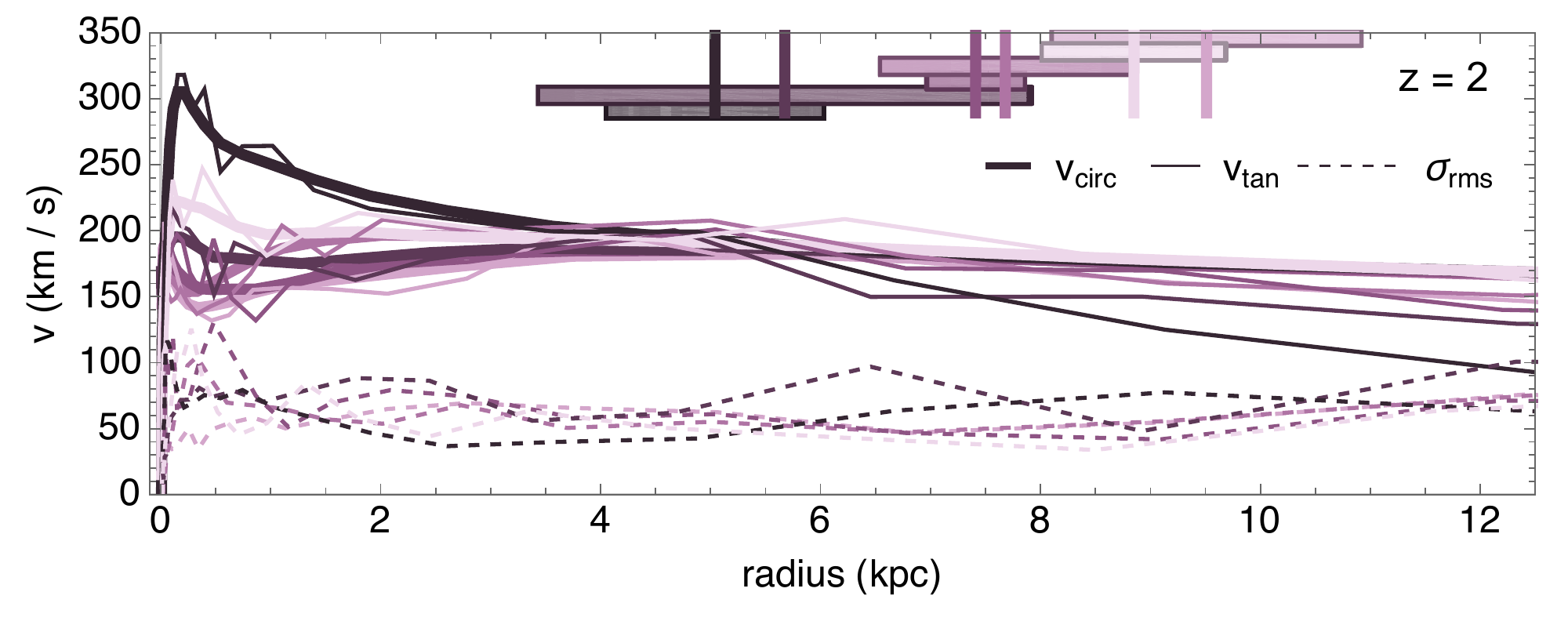}\\
	\caption{Rotation curves for the gaseous component of the galaxy at $z = 4$ and $z = 2$ for the inner galactic region ($r < 0.2\; r_\text{vir}$). Solid thick lines correspond to circular velocities $v_\text{circ} (r)$ (eq. \ref{eq:vCirc}), solid thin lines correspond to the tangential velocity $v_\text{tan} (r)$, and dashed lines to the turbulent velocity dispersion $\sigma_\text{rms} (r)$. Different colours display runs with increasing $B_0$: \textcolor{B20}{\textbf{MB20z}}, \textcolor{B14}{\textbf{MB14z}}, \textcolor{B13}{\textbf{MB13z}}, \textcolor{B12}{\textbf{MB12z}}, \textcolor{B11}{\textbf{MB11z}}, and \textcolor{B10}{\textbf{MB10z}} from the lightest to darkest purple. Vertical marks at the top of the plot panels correspond to $3 R_s$. Shaded bands around vertical marks correspond to errors associated with these values.}%
	\label{RotCurves}%
\end{figure*}

The observed reduction in the spin parameter of the gaseous component as $B_0$ increases implies a transfer of gas angular momentum. In the absence of enough rotational support, the galaxy shrinks radially, hence the measured decrease of $R_s$. To better quantify this, in Fig. \ref{RotCurves} we plot the rotation curves of the gas in the galaxy at $z = 4$ (top panel) and $z = 2$ (bottom panel). At $z = 4$, rotational support of the disk is not yet completely established (see Fig. \ref{GalDyns}, left column, third panel from the top). $\sigma_\text{rms} (r)$ closely follows the circular velocity of the gas $v_{\text{circ},\text{gas}}$ and shifts to the circular velocity of the dark matter $v_{\text{circ},\text{DM}}$ at approximately $3 - 4\,R_s$\footnote{The decomposition of the circular velocity is not shown in Fig. \ref{RotCurves} for sake of clarity}. Note that this is in approximate agreement with the radial distance at which the observed stellar density profile of local galaxies is found to be truncated \citep{Barteldrees94}. However, visual inspection of the galaxies does not highlight particular features at $3 - 4 R_s$, which argues in favour of a continuous transformation process rather than a sharp transition.

We plot on the bottom panel of Fig. \ref{RotCurves} rotation curves at $z = 2$, after  considerable shrinking of the galaxy {\it MB10z} (and to some extent, {\it MB11z}) has taken place. These rotation curves are in general agreement with observations of luminous $L_\star$ disk galaxies at $z = 2$. Our simulations match typical rotation velocities $v_\text{tan} \sim 150 - 200\; \text{km/s}$ \citep{Sofue01} and turbulent support 
$\sigma / v_\text{tan} \sim 0.1 - 0.3$, with $\sigma \sim 30 - 80\; \text{km/s}$ \citep{Erb04, Cresci09}. {\it MB10z} displays a much higher central peak velocity than the other runs, dominated by a more concentrated stellar component. 

At this redshift ($z \sim 2$, but also at $z \sim 4$), the larger $B_0$, the larger the deviation of $v_\text{tan}$ from $v_\text{circ}$ at distances $r \gtrsim 3\; R_s$. Rotation curves (thin solid curves) remain quite flat until the largest distances displayed in Figure \ref{RotCurves} ($r \sim 0.2 r_\text{vir}$; outer part of the galactic region) are reached, but display a steeper negative gradient as $B_0$ is increased. To a certain extent {\it MB11z}, but primarily {\it MB10z}, show decreasing rotation curves. This is enhanced for {\it MB10z} by the existence of a considerable central peak in $v_\text{tan}$.
Recently, \cite{Genzel17} reported decreasing galaxy rotation curves at $z=2$, which
they interpret as evidence for a lack of dark matter and an increased velocity dispersion support at large radii. 
This is clearly not the case in our simulations where the velocity dispersion support remains constant throughout the galaxy at $z=2$ and dark matter content is typical.
According to our findings, decreasing rotation curves could originate because of magnetic braking in the outskirts of galaxies, which naturally arises in MHD simulations of $\Lambda$CDM galaxies with a high amplitude of the primordial magnetic field ($B_0 \gtrsim 10^{-12} G$). 

The steepness of our rotation curves in the inner region also increases with the light concentration of the galaxies (see Fig. \ref{Colours}), in accordance with observations \citep{Swaters09}. This supports the picture that magnetic fields do not abruptly alter the evolution of spiral galaxies, but rather lead to higher central concentrations by gradually driving gas mass inward.

Magnetic braking in spiral galaxies can potentially operate through complementary channels. One possibility is the direct outward transport of angular momentum by toroidal Lorentz stresses. In this case, the radial and zenithal field lines are stretched azimuthally, and unbend at larger radii where the pressure is lower. Another possibility is radial and/or vertical deflection of the gas orbital trajectories. Finally, radial deceleration of inflows could reduce the supply of angular momentum, and inward magnetic acceleration of gas inside the galaxy by zenithal magnetic lines could lead to turbulent dissipation of galactic angular momentum.

\begin{figure*}%
\includegraphics[width=1.5\columnwidth]{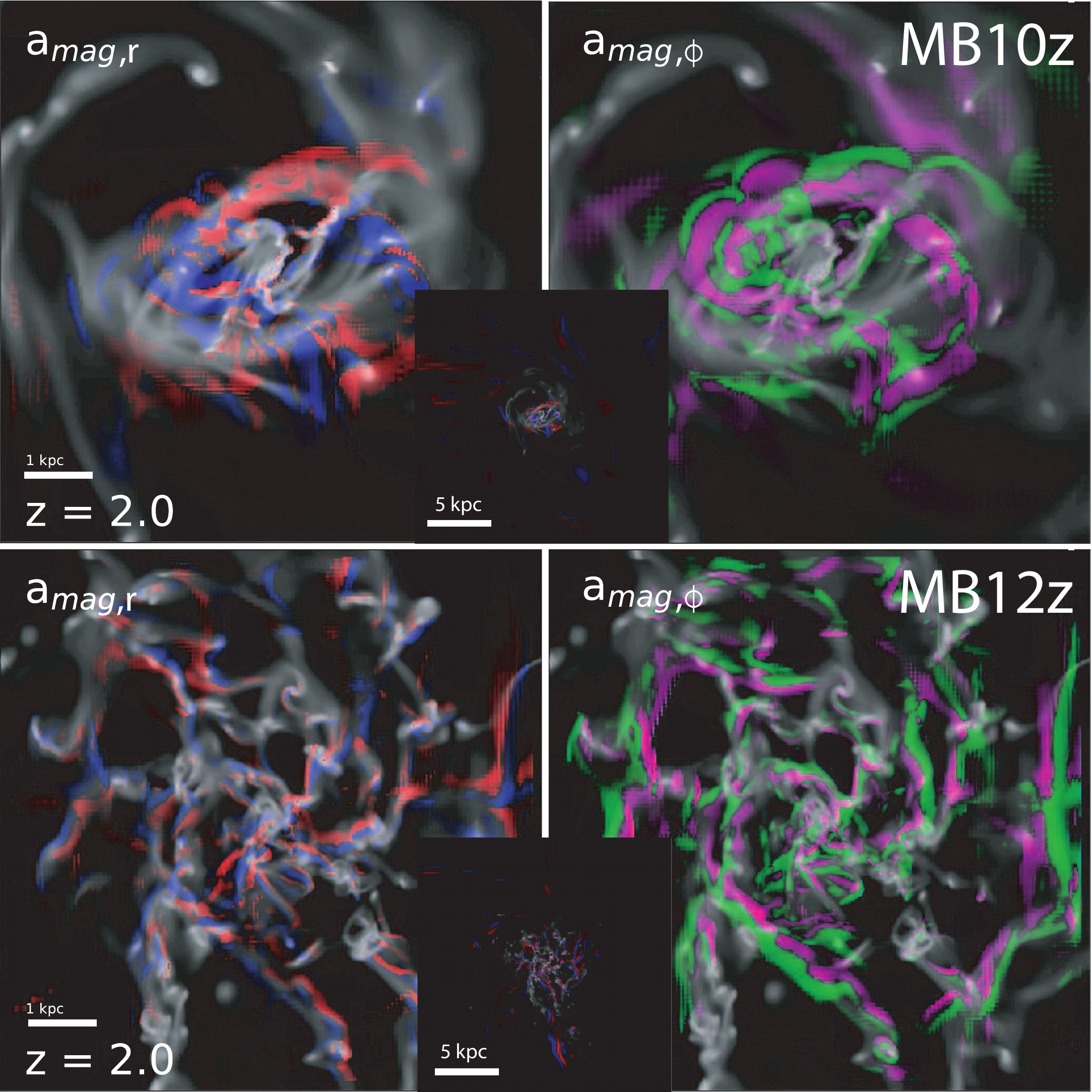}\\
\caption{Magnetic acceleration due to toroidal (left column) and radial forces (right column) for the {\it MB10z} (top row) and {\it MB12z} (bottom row) runs. Colours represent magnetic orbital braking (red), magnetic orbital acceleration (blue), inward magnetic acceleration (magenta), and outward acceleration (green). Gas density is overplotted using a grayscale. Regions of coherent magnetic acceleration are easily identified, and have larger sizes for higher $B_0$. In both runs magnetic orbital braking (red) dominates over magnetic orbital acceleration (blue). Magnetic stresses trigger stronger radial forces towards the centre of the galaxy, and higher outward pressure in the external parts (see also Fig. \ref{BrakeInflows}). The smaller inset panels, displaying a larger scale view, illustrate that toroidal magnetic forces are also important in the outskirts of the galaxy.}%
\label{BrakeImage}%
\end{figure*}

In order to quantify the impact of magnetic Lorentz force, we study how they affect gas orbits. We restrict our analysis to the most direct form of this force. We compute the radial and toroidal components of the Lorentz force $F_L$ on a disk of gas in a cylindrical coordinate frame $(r, \phi, z)$, where the $z$ dimension is aligned with the total angular momentum of the galaxy
\begin{equation}
    \begin{split}
    F_{L,r} = \vec{F_L} \cdot \hat{r} = \left[\left( \vec{B} \cdot \vec{\nabla} \right) \vec{B} - \frac{1}{2} \vec{\nabla} B^2\right] \cdot \hat{r} = \\
    B_r \stackrel{(1)}{\partial_r B_r} + \frac{B_\phi}{r} \stackrel{(2)}{\partial_\phi B_r} + B_z \stackrel{(3)}{\partial_z B_r} - \frac{1}{r}\stackrel{(4)}{B_\phi B_\phi} - \frac{1}{2 r} \stackrel{(5)}{\partial_r B^2},
    \end{split}
    \label{eq:LorentzFlow}
\end{equation}
\begin{equation}
    \begin{split}
    F_{L,\phi} = \vec{F_L} \cdot \hat{\phi} =  \left[\left( \vec{B} \cdot \vec{\nabla} \right) \vec{B} - \frac{1}{2} \vec{\nabla} B^2\right] \cdot \hat{\phi} = \\
    B_r \stackrel{(1)}{\partial_r B_\phi} + \frac{B_\phi}{r} \stackrel{(2)}{\partial_\phi B_\phi} + B_z \stackrel{(3)}{\partial_z B_\phi} + \frac{1}{r}\stackrel{(4)}{B_r B_\phi} - \frac{1}{2 r} \stackrel{(5)}{\partial_\phi B^2},
    \end{split}
    \label{eq:LorentzBraking}
\end{equation}
both expressed in rational units. 
We calculate these quantities term-by-term for a region centred on the galaxy as described in Section \ref{ss:GlobalProp}. Gradients are computed using central differences among neighbouring cells on the full AMR grid. 
In Fig. \ref{BrakeImage} we present close up views of the galaxy at $z=2$ which are density weighted maps of $a_{\text{mag},\phi} = F_{L,\phi} / \rho_g$, $a_\text{\text{mag},r} = F_{L,r} / \rho_g$ 
obtained by exclusively plotting forces in the galactic disk (i.e. $z < \left| 500\text{pc} \right|$) for two of our high magnetisation runs {\it MB12z} and {\it MB10z}. 
A positive contribution of $a_{\text{mag},\phi}$ leads to magnetic orbital acceleration (shown in blue), while a negative contribution leads to magnetic braking (shown in red).
When considering the overall spatial distribution of this toroidal acceleration, one can see that both magnetic orbital braking and acceleration occur primarily 
in dense gas structures extending all the way to the outskirts of the galaxy 
(as illustrated in the sub-panels of Fig.~\ref{BrakeImage}). 
Anti-symmetric acceleration and deceleration structures surround galaxy-scale magnetic field lines and magnetic pressure gradients. These regions coincide with gas spiral arms, which we find double as magnetic arms. This correlation is a behaviour frequently found in ideal MHD simulations \citep{Pakmor14, Mocz16, Butsky17}, which struggle to explain the observed displacement between magnetic and density spiral arms \citep{Chamandy15, Mulcahy17}. Interestingly, braking and inward forces marginally dominate magnetic forces in the dense gas in the inner parts of the galaxy ($r < 2$ kpc), especially for {\it MB10z}.

\begin{figure*}%
\includegraphics[width=1.92\columnwidth]{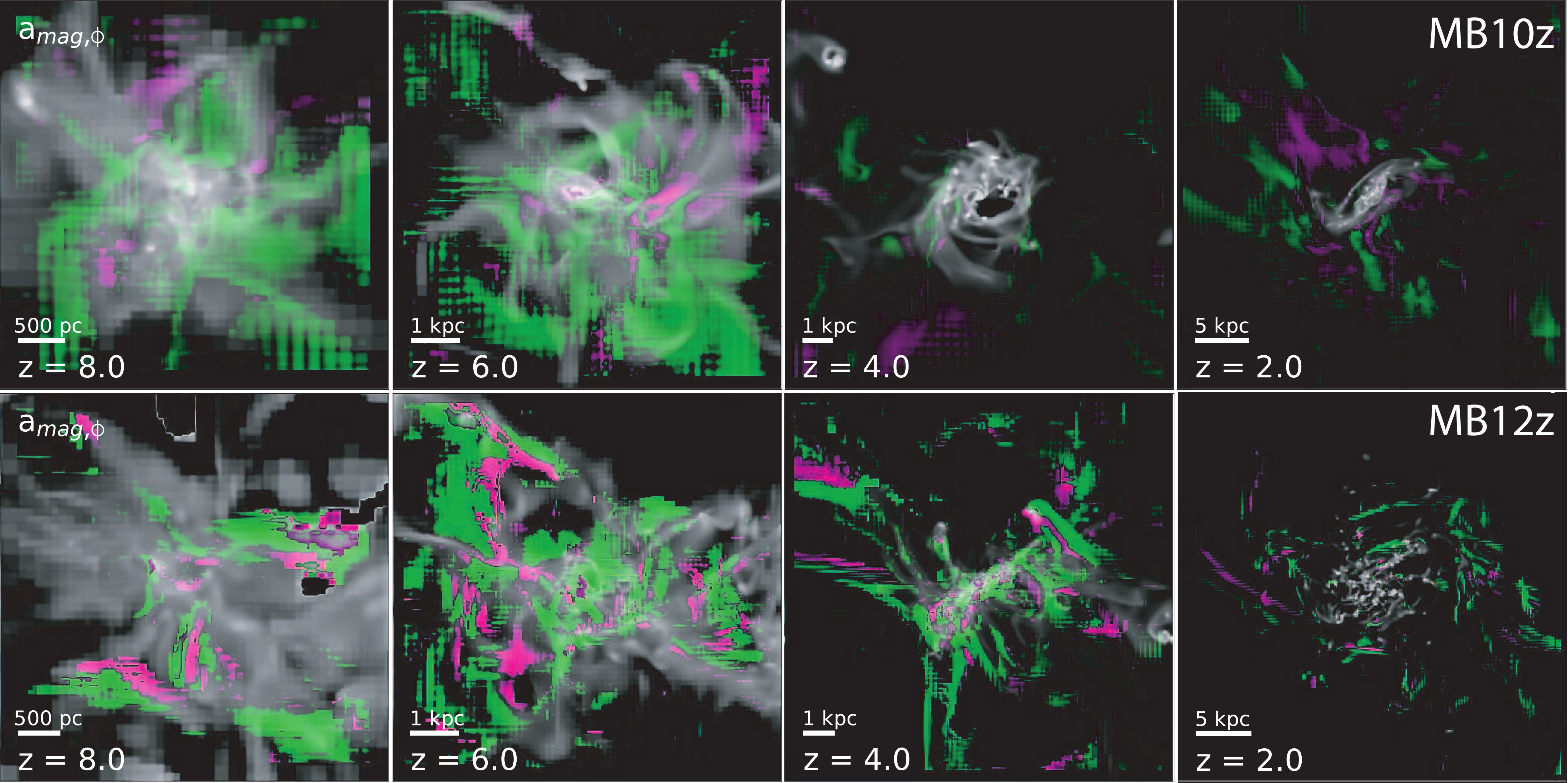}\\
\caption{(Top row) Radial acceleration for inflowing ($v_r < 0$) gas in the {\it MB10z} run. (Bottom row) Same as top row, but for the {\it MB12z} run. All project regions of $0.4\; r_\text{vir}$ on a side centred on the galaxy. Colour scales represent inward magnetic acceleration (magenta), and outward acceleration (green). Gas inflows appear to be dominated by magnetic forces working against accretion. As in Fig. \ref{BrakeImage}, gas density is overplotted using a grayscale. The dominance of the outward magnetic acceleration (green), largely a result of enhanced magnetic pressure (term (5) in eq. (\ref{eq:LorentzFlow})), could lead to a decrease of the angular momentum supplied to the galaxy (see text for detail).}%
\label{BrakeInflows}%
\end{figure*}

Inflows onto the galaxy also are subject to magnetic stresses, dominated by an outward force (with contributions from all terms in equation (\ref{eq:LorentzFlow})). We plot the radial magnetic forces on the inflowing gas (i.e. gas with $v_\text{gas,r} < 0$ in the frame of the galaxy) for the same two models ({\it MB12z} and {\it MB10z}) at various redshifts in Fig. \ref{BrakeInflows}. The dominance of the outward magnetic force (in green) over its inward counterpart (in magenta) leads to less angular momentum being supplied to the outskirts of the galaxy, facilitating inward gas transport. For lower $B_0$ values we find the size of the regions over which magnetic acceleration is coherent in Fig. \ref{BrakeImage} to be reduced. Large fluctuations in the magnetic pressure (term 5 in eqs. (\ref{eq:LorentzFlow}) and (\ref{eq:LorentzBraking})) dominate the magnetic acceleration throughout the galactic region. However, in our projections, these fluctuations of the magnetic pressure average out and the coloured structures observed in Fig. \ref{BrakeImage} are produced by a combination of terms (1), (2), (3), and (5) in eqs. (\ref{eq:LorentzFlow}) and (\ref{eq:LorentzBraking}). The density-weighted average magnetic acceleration is found to be on the order of $0.01$ to $1\; \text{km}\; \text{s}^{-1} \text{Myr}^{-1}$. However, higher values of coherent acceleration (or braking) appear locally, reaching up to several $10\; \text{km}\; \text{s}^{-1} \text{Myr}^{-1}$, both in {\it MB12z} and {\it MB10z}. We also find that the global magnetic acceleration roughly scales as $a_\text{Brake} \propto B_0$, becoming relevant for primordial magnetic fields $B_0 \gtrsim 10^{-13}$ G. The turbulence and stresses induced by opposing acceleration and braking components allow angular momentum to be transported outward. In the case of the radial part of the magnetic force, the inward component (shown in magenta) within the disk is predominant through contributions from terms (1), (3), (4). On the other hand, the outward component (shown in green) dominates in the outskirts, mostly thanks to the magnetic pressure contribution (term (5)). 

\begin{figure*}%
\includegraphics[width=\columnwidth]{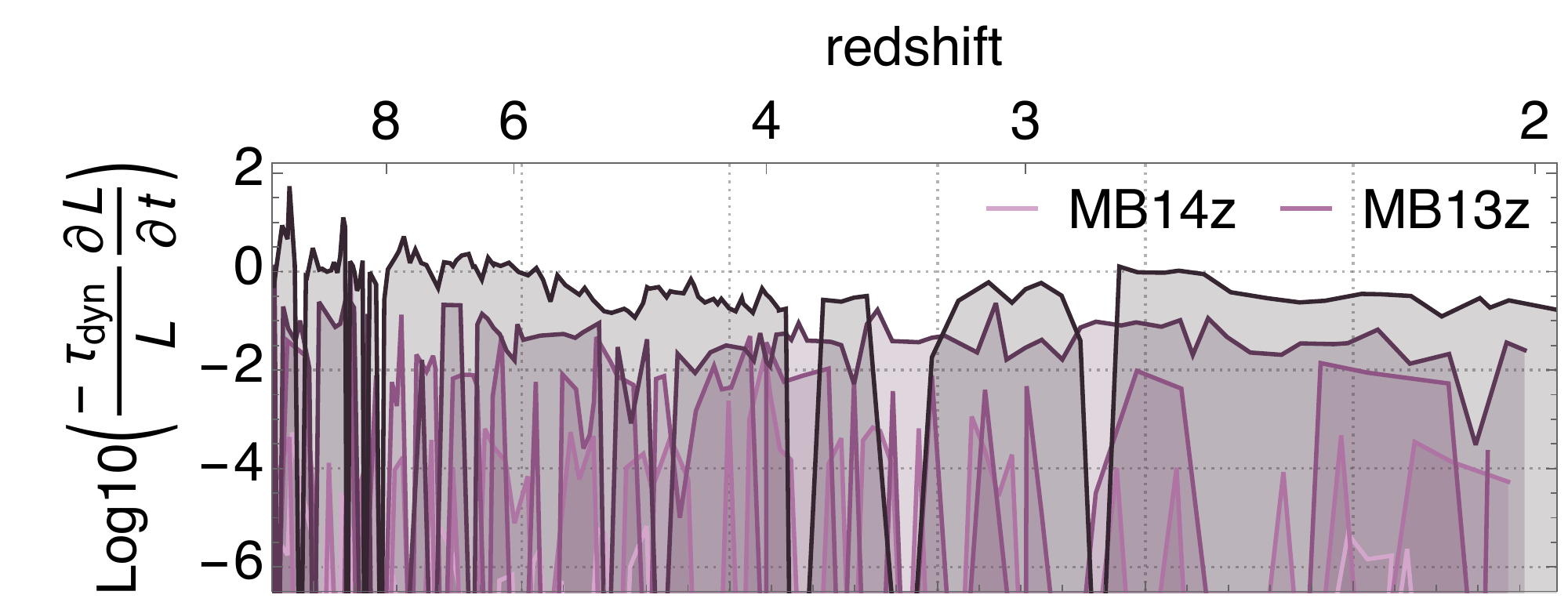}%
\includegraphics[width=\columnwidth]{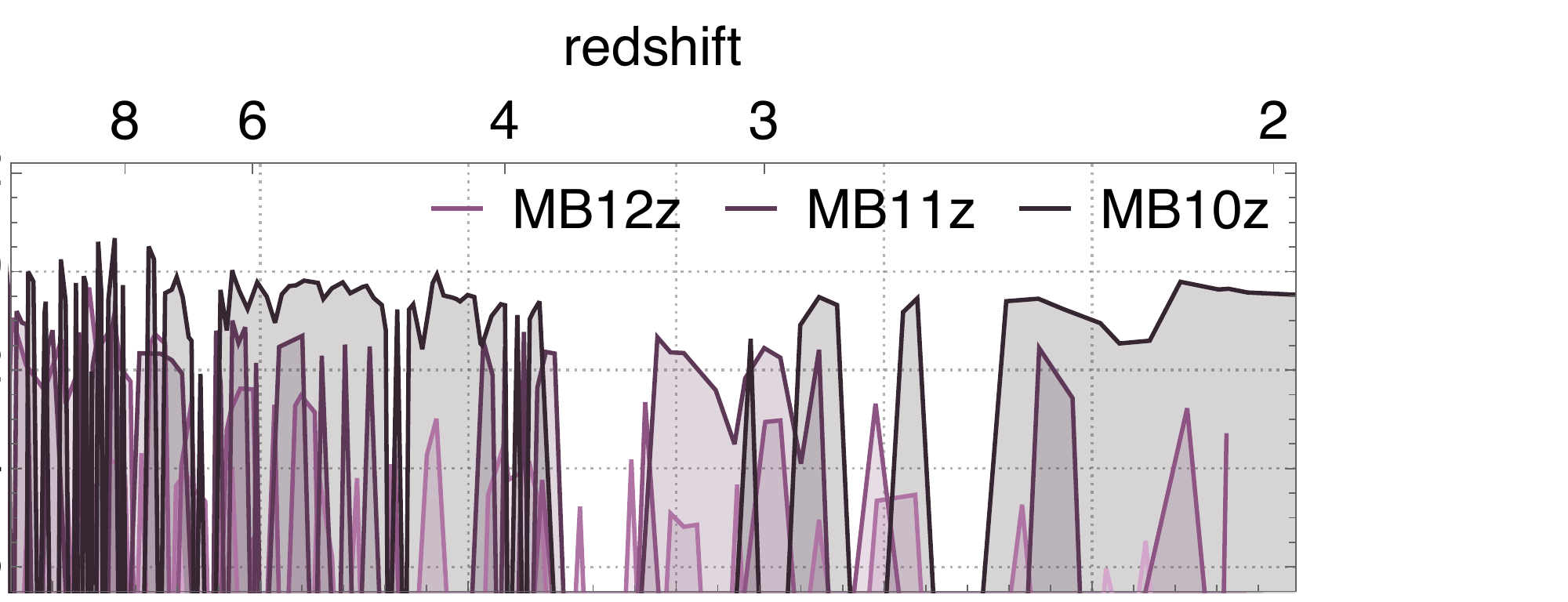}\\
\includegraphics[width=\columnwidth]{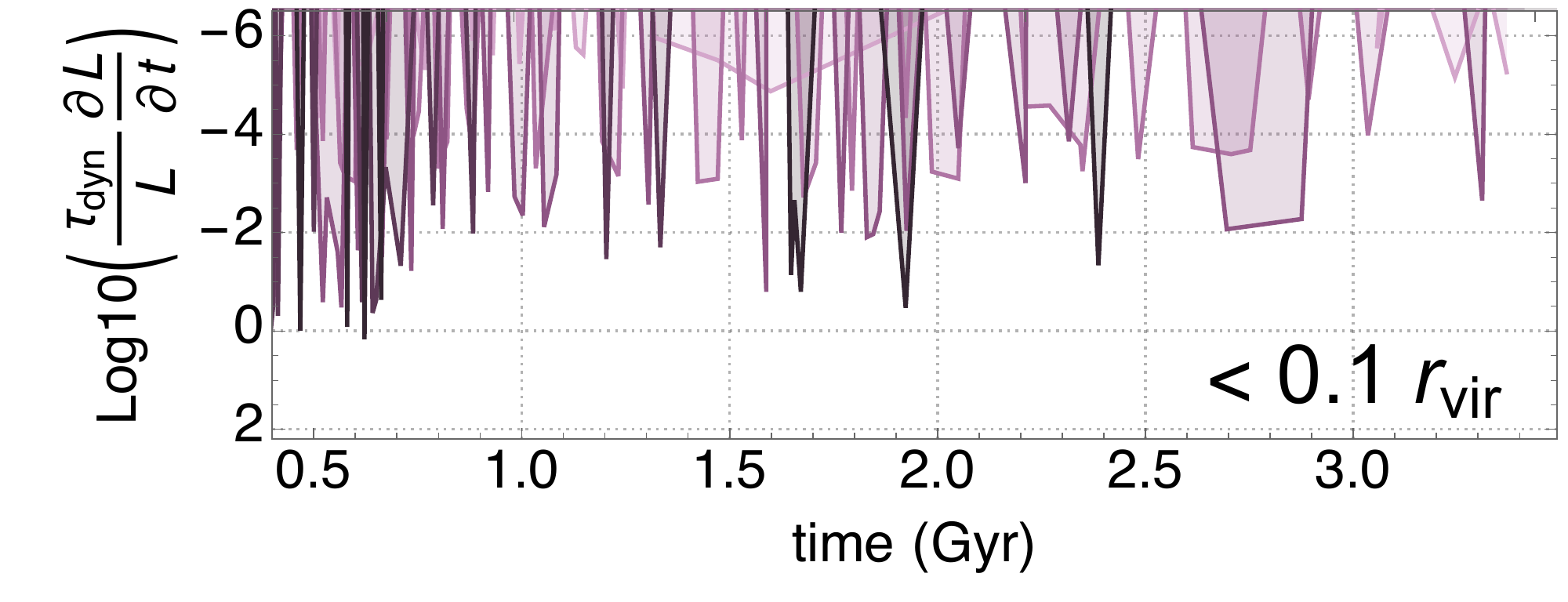}%
\includegraphics[width=\columnwidth]{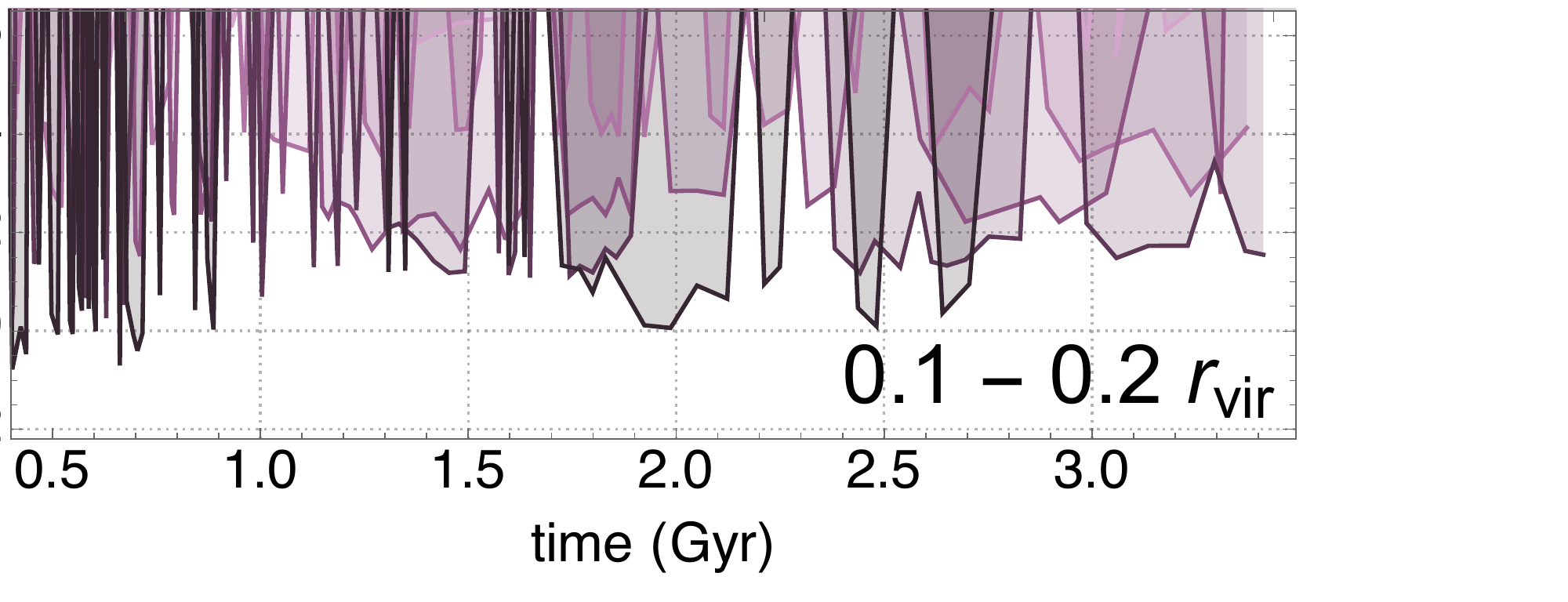}%
\caption{Spin-down parameters for the {\it MBz} runs. The left panels represent the inner fraction of the galactic region ($r / r_\text{vir}  < 0.1$), while the right panels correspond to its outer part ($0.1 < r / r_\text{vir} < 0.2$). Top panels correspond to relative angular momentum decrease per dynamical time, while bottom panels correspond to an increase of this quantity. The {\it MB10z} run clearly stands out, with strong magnetic fields affecting the dynamics of the galaxy (see text for detail).}%
\label{BrakeProfiles}%
\end{figure*}

To investigate more explicitly the dominance of braking forces (red coloured regions) in the inner and outer galactic region, and to study their dynamical importance, we perform the following estimate of the magnetic spin-down of the galactic angular momentum $L_z$ in the inner galactic region ($r < 0.1\; r_\text{vir}$)
\begin{equation}
    S_D^\text{in} = \frac{\partial_t L_z}{L_z} \tau_\text{dyn} =  \frac{\sum_{i = 0 r_\text{vir}}^{0.1 r_\text{vir}}{r_i\; \rho_{g,i}\; V_\text{cell, i}\; a_{\phi, i}}}{ \sum_{i = 0 r_\text{vir}}^{0.1 r_\text{vir}}{r_i\; \rho_{g,i}\; V_\text{cell, i}\; v_{\phi, i}}  } \tau_\text{dyn},
    \label{eq:SpindownIn}
\end{equation}
and the outer galactic region ($0.1 r_\text{vir} < r < 0.2 r_\text{vir}$)
\begin{equation}
    S_D^\text{out} = \frac{\partial_t L_z}{L_z} \tau_\text{dyn} =  \frac{\sum_{i = 0.1 r_\text{vir}}^{0.2 r_\text{vir}}{r_i\; \rho_{g,i}\; V_\text{cell, i}\; a_{\phi, i}}}{ \sum_{i = 0.1 r_\text{vir}}^{0.2 r_\text{vir}}{r_i\; \rho_{g,i}\; V_\text{cell, i}\; v_{\phi, i}}  } \tau_\text{dyn},
    \label{eq:SpindownOut}
\end{equation}
where we sum over all AMR grid cells inside the region. In these expressions, $r_i$ is the distance between a cell $i$ and the centre of the region, $\rho_{g,i}$ is the gas density of the cell, $V_\text{cell, i}$ corresponds to its volume, $v_{\phi,i}$ is the toroidal velocity of the gas in the cell around the galactic rotation axis in the frame of the galaxy (i.e. removing the bulk motion of the galaxy), and $\tau_\text{dyn}$ is the dynamical time of the galaxy at a given time computed as indicated in Section \ref{ss:GlobalProp}. The measured spin-down parameters correspond exclusively to direct magnetic stresses and therefore fail to capture the contribution to the change in galactic angular momentum from outflows, inflows, and other forces of a different nature. Bearing these caveats in mind, we show in Fig. \ref{BrakeProfiles} the resulting spin-down parameters (inner region in the left column panels, outer region in the right column ones).

The figure indicates that for the {\it MB12z}, the {\it MB11z} and especially the {\it MB10z} runs, magnetic braking is a non-negligible effect when considering the evolution of the angular momentum of the galaxy. It is also interesting to note that magnetic acceleration takes place during a significant fraction of time in the outskirts of the galaxy for all the magnetised {\it MBz} runs apart from {\it MB14z}. This could boost orbital velocities at large distances through a different mechanism to that proposed by \citet{Ruiz-Granados10}, i.e. a rise of the circular velocity produced by the radial decrease of toroidal magnetic field strength. However, we remark that for the galaxy mass and physical distances studied here, we found that the presence of magnetic fields decreases rather than increases orbital velocities. The importance of direct magnetic braking in {\it MB10z} and perhaps in {\it MB11z}, very likely explains the marked deviation of {\it MB10z} from the other runs. The described morphological impact of magnetic fields appears to be more noticeable in the presence of stellar feedback. \citet{Shukurov2018} claim that the presence of strong magnetic fields in the ISM may reduce the speed of galactic outflows, thereby quenching them. This could cause higher $B_0$ values to reduce the capacity of stellar feedback to expel low angular momentum gas from the galaxy.

To summarise, we find evidence in our simulations for a mechanism that brakes the rotation of galaxies. Magnetic stresses appear to drive baryonic mass towards the centre of the galaxy, both through outward transport of galactic angular momentum and magnetic deceleration of inflowing gas. These processes reduce the angular momentum accreted by the galaxy by accelerating inflows outwards by means of a magnetic pressure dominated Lorentz force (term (5) in eq. (\ref{eq:LorentzFlow})). Note that this magnetic pressure gradient is expected to be present to some degree even when the magnetisation is not of primordial origin. Other possible consequences of these strong magnetic fields are put forward by \citet{Sethi10} and \citet{Pandey19}, where the former authors argue that heating from a primordial magnetic field  $B_0 \sim 4 \cdot 10^{-9}$ G  could avoid fragmentation of primordial gas and lead to direct collapse onto super massive black holes, while the latter authors discuss how strong primordial magnetic fields of $B_0 \sim 10^{-10}$ G can dramatically reduce the angular momentum of infalling gas into massive haloes through tidal torques. According to their calculations, this angular momentum loss also facilitates the formation of direct collapse black holes. 

 Finally, given the intensity of the effects that we measure, strong magnetic fields can also contribute to the formation of bulges through magnetic forces and magnetic braking (see Fig. \ref{GalDyns}, bottom right panel). Although this is not explored in this work, both concentration parameters (Fig. \ref{Colours}) and the centrally peaked circular velocity of our {\it MB10z} run hint at the potential role played by magnetic fields during galaxy mergers \citep{Wang09}. Magnetic braking might also be at play in the outskirts of galaxies magnetised through mechanisms other than large-scale seeding (e.g. astrophysical sources). These other mechanisms are expected to provide magnetic energy to thermal energy ratios comparable to the {\it MB12z} or {\it MB11z} runs. 
 
\subsection{Magnetic fields and galactic observables at z = 2}
\label{ss:LowZ}

\begin{figure*}%
\includegraphics[width=2\columnwidth]{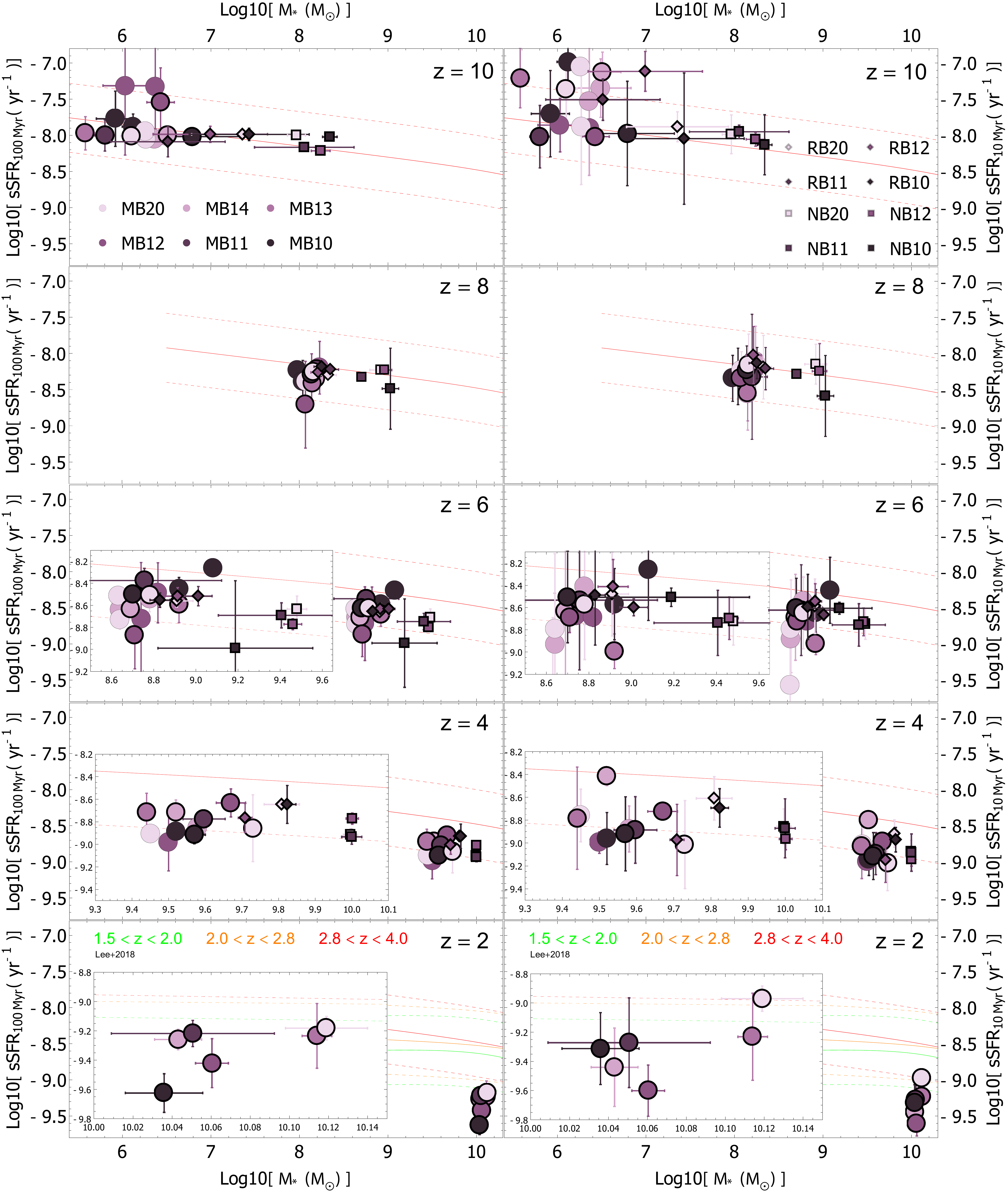}
\caption{Specific star formation rate (sSFR) vs stellar mass ($\Mst$) for different strength of the primordial magnetic field $B_0$. The larger the strength of the magnetic field the darker the shade of purple of the corresponding data points. The left column corresponds to a sSFR averaged over a 100 Myr period, while the right column represents sSFR averaged over 10 Myr. Data points legends are the same as for Figs. \ref{GalScales} and \ref{GalDyns}. Subsequent rows correspond to $z =$ 10, 8, 6, 4, and 2 respectively from top to bottom. Subpanels are zooms around the bulk of the distribution of data points. Solid coloured lines correspond to the sSFR - $\Mst$ main sequence (MS) relation estimate obtained by \citet{Lee18} from their observations in redshift intervals $2.8 < z < 4.0$ (red), $2.0 < z < 2.8$ (orange), and $1.5 < z < 2.0$ (green) respectively. Dashed lines present factors of three above and below the solid line estimates. In the two columns of the upper panels ($z > 4$), higher values of $B_0$ place our galaxy closer to the extension of the MS obtained by \citep{Lee18} for $2.8 < z < 4.0$ to higher redshifts. Contrarily, at $z = 2$, lower $B_0$ appears to situate the galaxy closer to the center of this MS. Overall, the studied range of primordial magnetic fields do not seem to display any clear systematic effect on either sSFR-$\Mst$ or the final $\Mst$ of our galaxy, in agreement with previous studies \citep[see text;][]{Su17}.}%
\label{GalStar}%
\end{figure*}

Having demonstrated that strong magnetic fields modify global properties of galaxies, in this section we review whether such an impact could leave an observational signature detectable with upcoming facilities such as {\it JWST}.

The stellar mass $\Mst$ is one of the most fundamental properties of a galaxy, and represents the integrated star formation over time. Star formation rates are regulated by gas accretion \citep[e.g][]{SanchezAlmeida17} and stellar feedback \citep[e.g][]{Hayward17}. Both the stellar mass of a galaxy and its star formation rate (SFR) can be measured by observations of galaxies using various estimators \citep{Kennicutt12}. In Fig. \ref{GalStar} we present the changes in specific star formation rates (sSFR) vs stellar masses in the galactic region when the strength of the primordial magnetic field $B_0$ varies. This sSFR takes into account the amount of stars with ages $t - t_\text{birth}$ less than a timescale $\Delta t$
\begin{equation}
    \text{sSFR}_{\Delta t} (t) = \frac{\text{SFR}_{\Delta t}}{\Mst} (t) = \frac{1}{\Mst} \frac{M_* \left(0 \leq t - t_\text{birth} < \Delta t\right)}{\Delta t},
    \label{eq:sSFR}
\end{equation}
where $t_\text{birth}$ is the time of formation (birth) of a star particle. We study the sSFR averaged over the entire galactic region employing different timescales $\Delta t$: a longer one consistent with estimators such as UV or FIR ($\Delta t = 100$ Myr; Fig. \ref{GalStar}, left column), and a shorter one, more frequently associated with $H\alpha$ observations ($\Delta t = 10$ Myr; Fig. \ref{GalStar}, right column). We measure for each simulation output $\text{SFR}_{\Delta t}$ and $M_*$ in the galactic region. However, for Fig. \ref{GalStar}, we compute the values for $M_*$ (horizontal axis) and $\text{sSFR}_{\Delta t}$ (vertical axis) for the outputs around the target redshift following the process described in Section \ref{ss:Measurements}.

\begin{figure}%
	\includegraphics[width=\columnwidth]{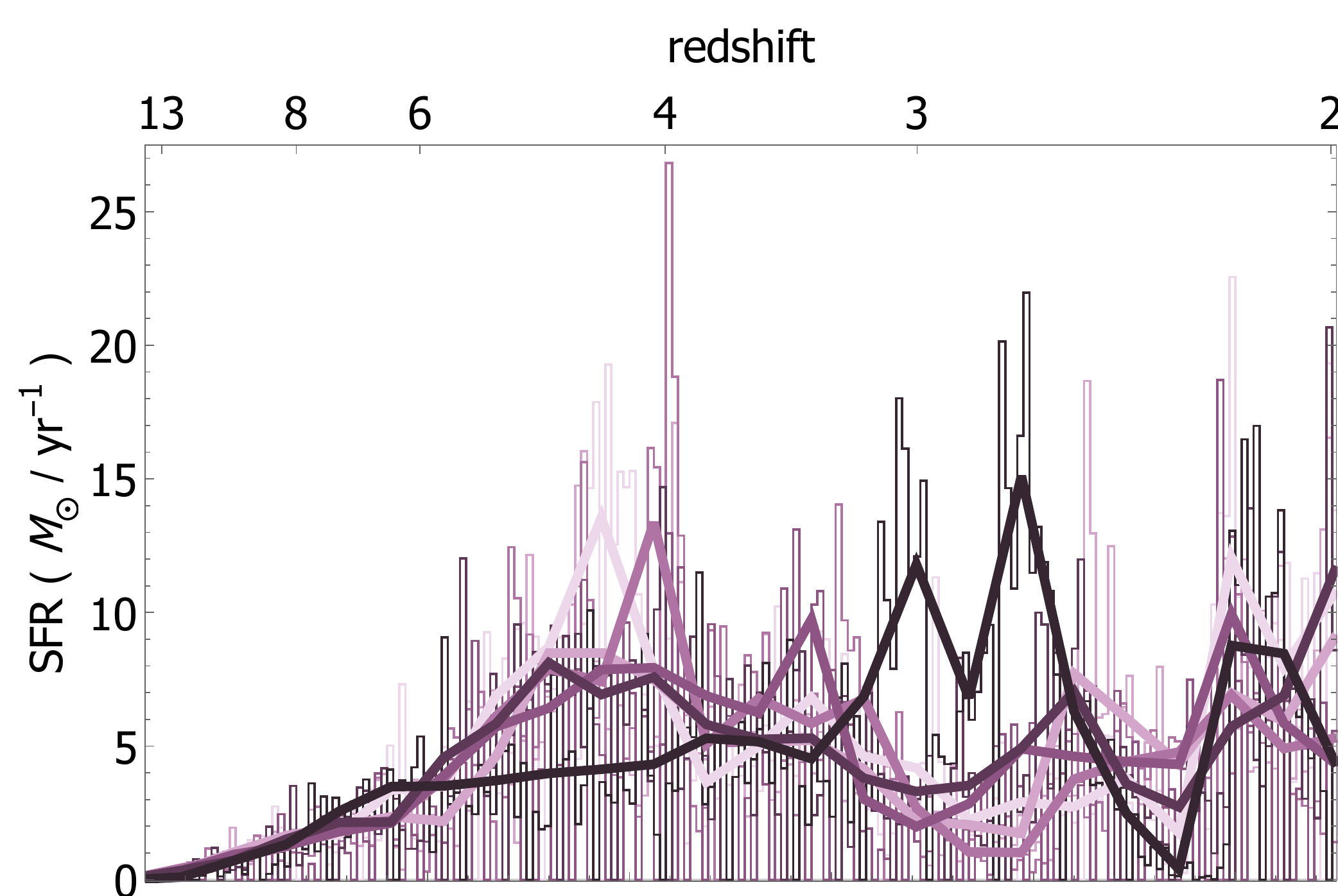}
	\includegraphics[width=\columnwidth]{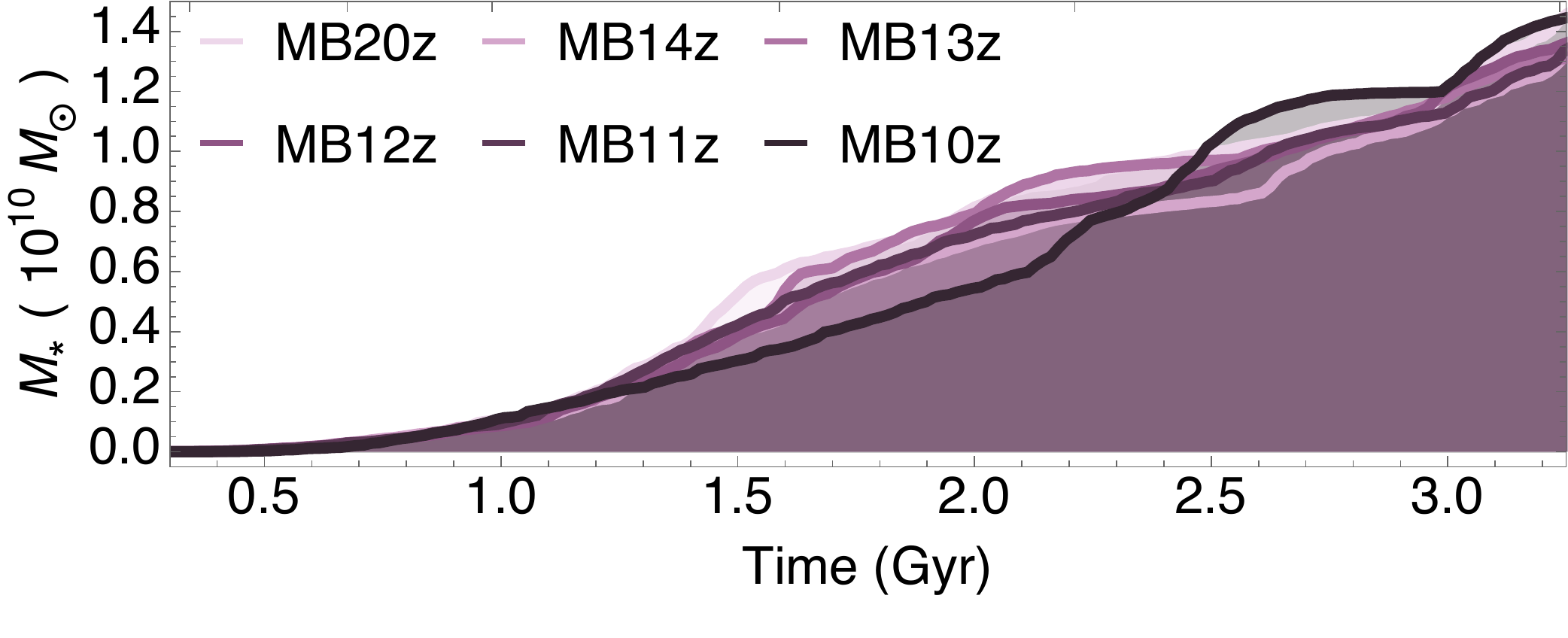}
	\caption{(Top) Star formation history for each of the {\it MBz} runs as a function of time, average over shorter ($\sim 10$ Myr; thin bins) and longer ($\sim 100$ Myr; thick lines) periods of time. (Bottom) Cumulative stellar mass in the galactic region for each of the runs, corresponding to the integral of the star formation rate plotted in the top panel. As for previous figures, darker shades of purple indicate higher $B_0$. Stronger $B_0$ delays the growth of the cumulative stellar mass, yet all simulations end with approximately the same stellar mass by $z=2$.}%
	\label{SFH}%
\end{figure}

While magnetic fields are expected to play a major role in the process of star formation on small scales, Figs. \ref{GalStar} and \ref{SFH} show that they have an insignificant impact on the final stellar mass of galaxies, in accordance with previous studies \citep{Su17}. As expected, we find that the most important factor at play in the evolution of $\Mst$ is the stellar feedback prescription employed. Varying it introduces variations of $\sim 0.2 - 0.5$ dex, while removing the feedback altogether can boost stellar masses by up to an order of magnitude. As redshift decreases, the two feedback prescriptions employed converge in terms of $\Mst$. From the highest redshift down to $z = 2$, it is hard to establish any systematic effects due to the presence of magnetic fields. The stellar mass of the galaxy appears as insensitive to the orientation of the primordial magnetic field as to its strength, which seems to legitimate the standard usage of uniform (or even lack of) primordial magnetic fields for large-scale simulations that aim to produce stellar mass functions, at least for galaxies with masses $\sim M_\star$ as we consider in this work. We stress that the presence of magnetic fields alters other global galaxy properties, like their sizes, as we have seen, but not their stellar mass. Whether the final stellar mass would remain unchanged by magnetic fields if the process of star formation was better captured (e.g. by employing higher spatial resolution and/or including better sub-grid models) is not clear. While the panels of Fig. \ref{GalStar} show that there is a non-negligible spread in the SFR for both indicators, magnetic fields do not seem to have a systematic effect with SFR curves criss-crossing one another, regardless of the feedback implementation. We find larger deviations for the 10 Myr measurement (right hand panels in Fig. \ref{GalStar}), but this is expected.

Magnetic fields driving gas inwards in the galaxy should cause a stronger depletion of gas in the inner regions of the galaxy through star formation, and thus the sSFR should decrease further at later times when stronger magnetic fields are present. Looking at the bottom panels of Fig. \ref{GalStar}, the evidence based on our 6 simulations is at best circumstantial. We therefore conclude that primordial magnetic fields do not seem to have a systematic effect on the global sSFR of our galaxies during the period studied.

We include in Fig. \ref{GalStar} coloured lines corresponding to the sSFR - stellar mass main sequence relation (MS) of star formation obtained by \citet{Lee18}. Solid lines represent the slope estimates provided by these authors for the SFR vs $\Mst$ relation using their equation (4). Dashed lines correspond to estimates a factor of 3 above and below. We show in red, orange and green the fits corresponding respectively to the $2.8 < z < 4.0$, $2.0 < z < 2.8$, and $1.5 < z < 2.0$ redshift binning of their data. Due to the absence of matching data, we compare their results for $2.8 < z < 4.0$ to our higher redshift measurements ($z > 4$). We find that our runs with higher $B_0$ seem in better agreement with the extrapolated line. Contrarily, at $z = 2$, we find that weaker $B_0$ galaxies lie closer to the observed MS relationship overall. While our galaxies appear to lie systematically below this relation at $z \leq 4$, presumably as a result of not boosting the stellar feedback prescription, the separation of the data points with respect to the MS does not change with varying primordial magnetic field strengths. Therefore, we also do not find conclusive evidence that primordial magnetic fields drive the galaxies in our simulations towards or away from the star formation MS. Magnetic fields do nonetheless alter the time evolution of the stellar mass and spatial distribution of star formation. Regarding the time evolution, while stellar masses in all simulations become remarkably similar with time (by $z = 2$, their relative variation is $\Delta \Mst / \Mst \sim 1\%$), the star formation history (SFH) of the galaxy differs for runs with different primordial magnetic field strengths. We show this in Fig. \ref{SFH}, where we display the SFH for each of the {\it MBz} runs and their time integral, namely, their cumulative stellar mass.

As the strength of $B_0$ increases in Fig. \ref{SFH}, high SFRs are shifted towards later epochs, pushing the peak of star formation around $t \sim 1.5$ Gyr ($z \sim 4$) to $t \sim 2.3$ Gyr ($z \lesssim 3$). The deviations become significant for $B_0 > 10^{-12}$ G ({\it MB12z} and {\it MB10z}). Such a behaviour (delay of the onset of star formation) is also reported in MHD studies of SFR on molecular cloud scales \citep{Hennebelle14}. At approximately $t \sim 2.3$ Gyr, all simulations display a secondary peak of star formation, associated with a merger. The strength of this peak is slightly increased as $B_0$ increases. For {\it MB10z}, we find an extended period of high star formation at $z \lesssim 3$, during which the cumulative stellar mass catches up with the simulations featuring lower $B_0$ strengths.

\begin{figure}%
	\includegraphics[width=\columnwidth]{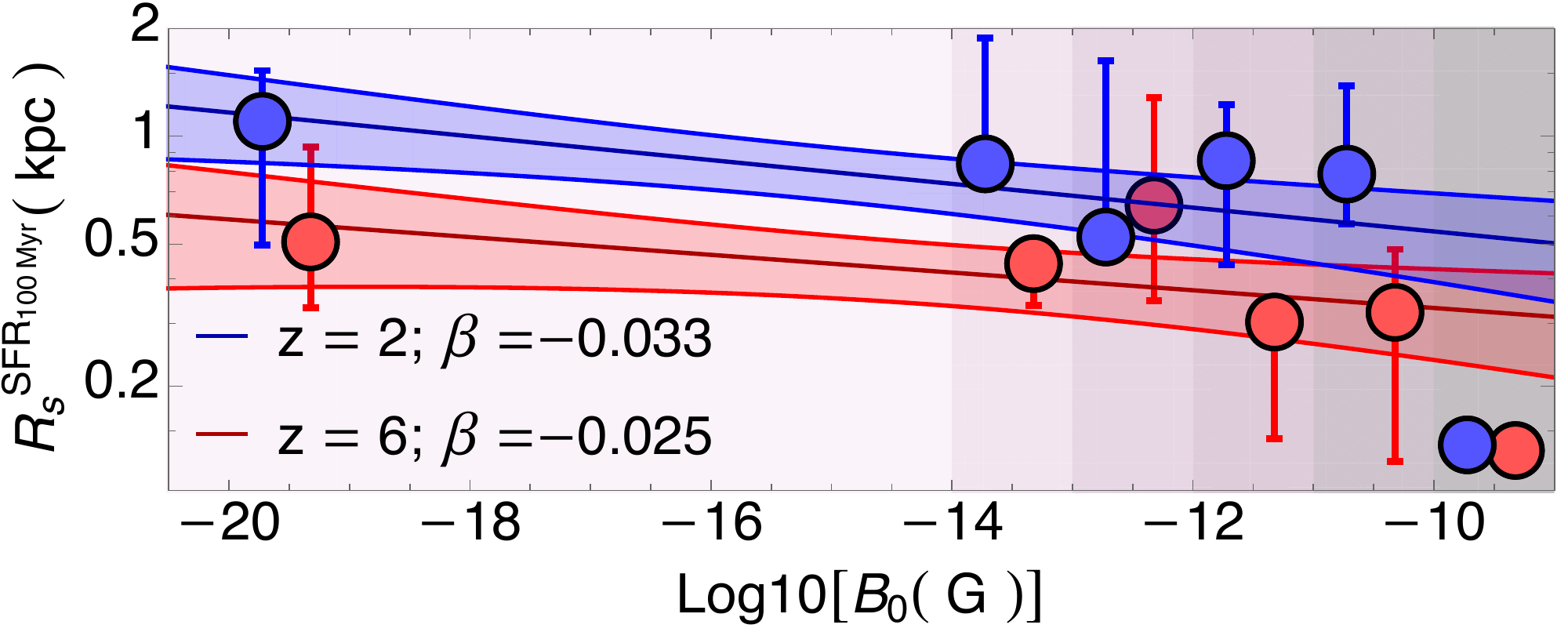}\\
	\caption{Changes in the galaxy $\text{SFR}_{100 \text{Myr}}$ radial scale length $R_s^{\text{SFR}_{100 \text{Myr}}}$ as a function of $B_0$ during the accretion phase ($z = 6$) and the feedback phase ($z = 2$) for the {\it MBz} runs. Data points include minor displacements in the x-axis to improve readability. Error bars as for Fig. \ref{GalScales}. Solid lines correspond to data fits to $R_s^{\text{SFR}_{100 \text{Myr}}} = \alpha {B_0}^\beta$ and 68\% confidence intervals. We find the star formation to be slightly more concentrated towards the centre of the galaxy as $B_0$ increases.}%
	\label{SFRscales}%
\end{figure}

On top of modifying the SFH, the presence of magnetic fields also affects the spatial distribution of star formation and the stellar component, and to some degree $\sigma_\text{rms}^\text{stars}$. We briefly explore the changes in the distribution of star formation in Fig. \ref{SFRscales}. It shows the $\text{SFR}_{100 \text{Myr}}$ (as described by equation (\ref{eq:sSFR})) radial scale length $R_s^{\text{SFR}_{100 \text{Myr}}}$, computed according to the method described in Section \ref{sss:Morphology}. For $B_0 < 10^{-10}$ G, we observe a modest trend for the star formation distribution to be slightly more concentrated towards the centre of the galaxy as $B_0$ is increased. This occurs both at very high redshift (accretion phase, $z = 6$) and at the lowest redshift studied (feedback phase, $z = 2$). We fit the radial scales at each redshift to the function $\alpha {B_0}^\beta$ (Fig. \ref{SFRscales}, solid lines) to better confirm this weak scaling. However, for the {\it MB10z} run ($B_0 > 10^{-10}$ G) we find a considerable concentration of the star formation towards the centre of the galaxy.

\begin{figure*}%
	\includegraphics[width=\columnwidth]{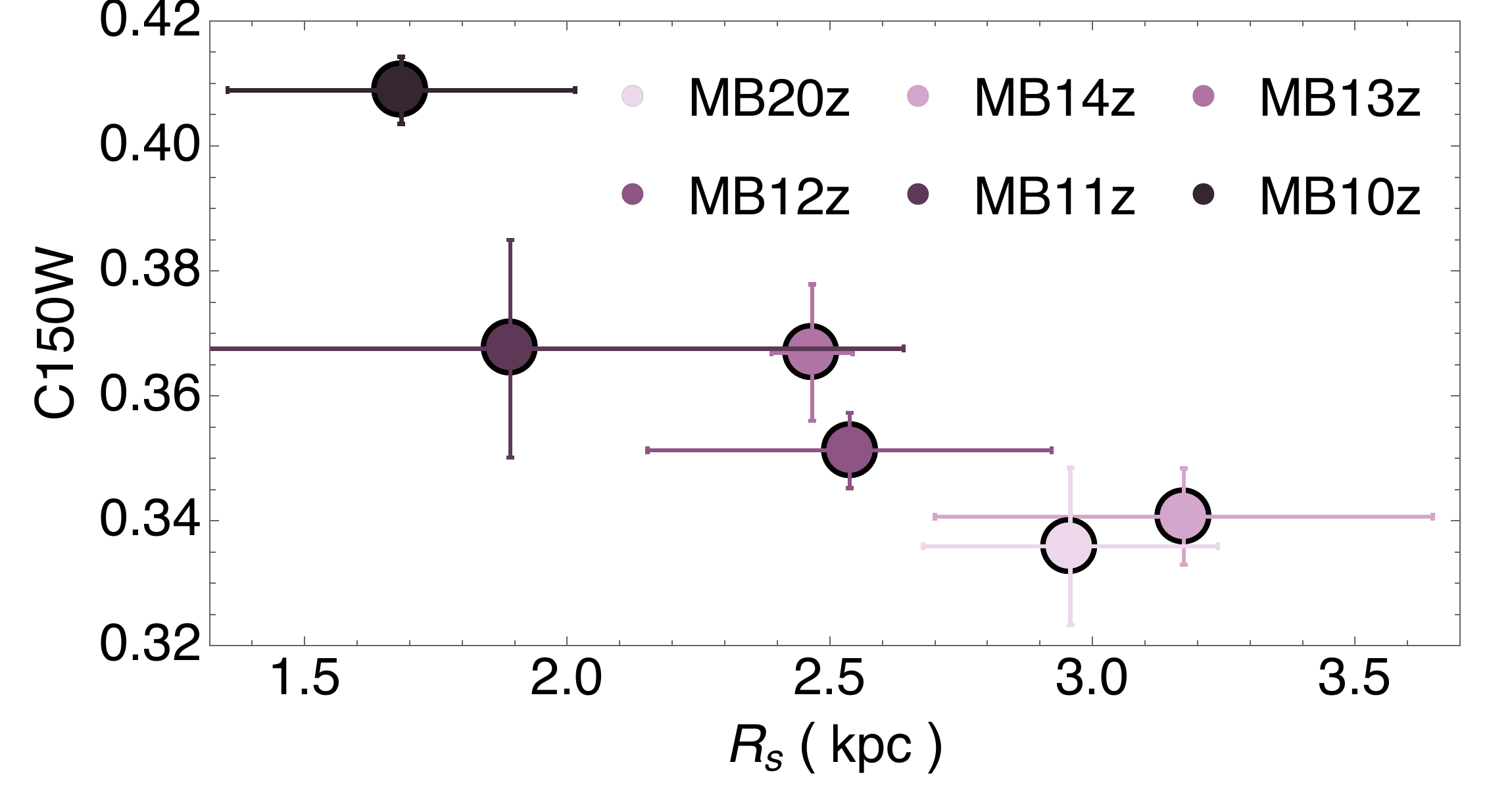}
	\includegraphics[width=\columnwidth]{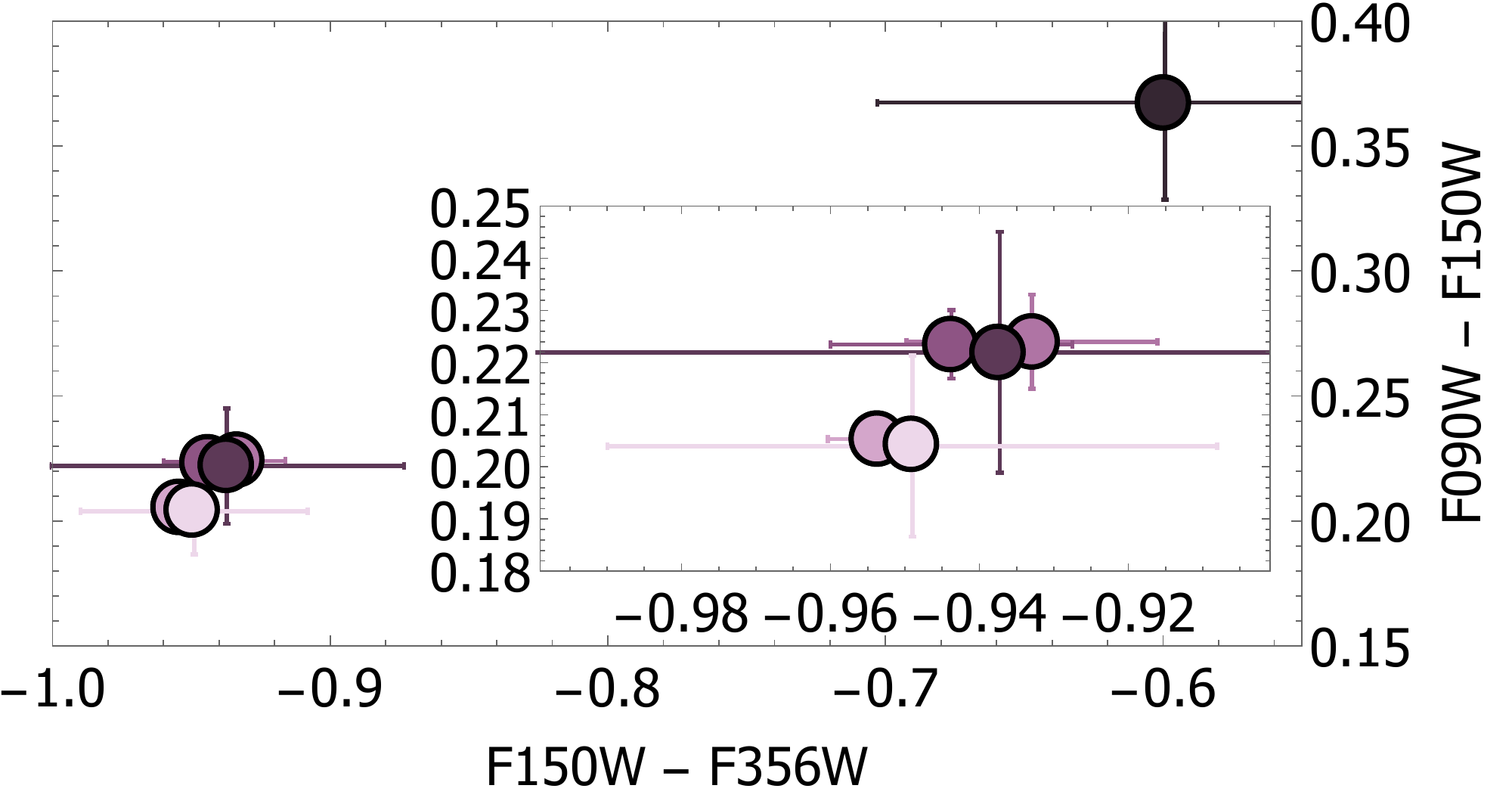}\\
	\caption{Variation of the rest-frame $V$-band concentration parameter (left; $C150W$) and rest-frame $U - V$ vs $V - J$ colours (right; $F090W - F150W$ vs. $F150W - F356W$) for various strengths of the primordial magnetic field at $z = 2$. See Section \ref{ss:GlobalProp} for details on the procedure to calculate these quantities. An increase of $B_0$ leads to a systematic reduction of the gas radial scale length of the galaxy and an increase in its $V$ concentration parameter.}%
	\label{Colours}%
\end{figure*}

The variations found for the properties of the galaxy at high $B_0$ values should in principle be reflected in observational parameters such as the colour or the concentration of galaxies. Observations of massive galaxies find that strong starbursts occur in very compact nuclear regions at $z \gtrsim 4$, leading very rapidly to high stellar masses and build up of concentrated stellar cores \citep[e.g.,][]{Toft14, Ikarashi15, Oteo17, Gomez-Guijarro18}. As their stellar population result from a high redshift ($z \gtrsim 4$) burst, the compact stellar cores should display redder colours by $z = 2$. Measuring stellar concentrations and colours as a function of magnetic field strength in our simulations could thus shed light on the potential influence of magnetic fields on the SFH of galaxies. In particular, both higher stellar concentration and redder colours could distinguish bursty SFHs from smoother secular evolution. the rest-frame $UVJ$ bands are of particular interest because they are the wavebands where the most prominent changes in spectral features occur as the stellar populations evolve \citep{Patel12}. Indeed, the rest-frame $U - V$ and $V - J$ colours have been widely used to identify and study post-starburst galaxies \citep[e.g.][]{Wild14, Wilkinson17}.

We thus average colour and concentration measures over $\tau_\text{dyn} = 0.4$ Gyr to generate mock {\it JWST} observations at $z = 2$ (as described in Section \ref{ss:Sunset}). Fig. \ref{GalaxiesZ2} presents face-on rest-frame {\it UVJ} snapshots of the studied galaxy from the runs available at that redshift ({\it MB20z}, {\it MB14z}, {\it MB13z}, {\it MB12z}, {\it MB11z}, and {\it MB10z}). These are obtained by convolving the galaxy spectrum with the [$F090W$, $F150W$, $F356W$] {\it JWST} NIRCam filters.
The rest-frame $V$-band concentration parameter $C150W$ (left), and rest-frame $U - V$ vs $V - J$ colours from $F090W - F150W$ vs $F150W - F356W$ (right) are shown in Fig. \ref{Colours}. An increase in $C150W$ (left panel) as the magnetic field increases is in agreement with the behaviour of $R_s$ and $R_s^\text{stars}$ previously discussed. 
On the other hand, the colours (right panel) are relatively independent of $B_0$, with a small colour reddening as $B_0$ increases; except the extreme {\it MB10z} run, where the galaxy becomes considerably redder, as expected after a major starburst event at $z \sim 3$. This observed redder colour arises as a lower fraction of stars form for {\it MB10z} over the past $\sim 0.5$ Gyr ($2.6\; \text{Gyr} \lesssim t \lesssim 3.2 \; \text{Gyr}$), i.e. in the aftermath of the major star formation burst, than for the other runs. Moreover, ongoing SFR during the measurement time interval provides a significant contribution to the measured colours. 

Finally, if this reduction in size and the ensuing concentration of light, which become more pronounced as magnetisation increases, also hold for dwarf galaxies, magnetic fields could be a key factor in the formation of the most compact dwarf systems reported by Local Group simulations \citep{Garrison-Kimmel2019}. We will investigate this in future work. 

\section{Conclusions}
\label{s:Conclusions}

In this manuscript, we employed high resolution cosmic zoom-in MHD simulations of a Milky Way-like spiral galaxy to explore the impact of changing the primordial magnetic field configuration on the global morphological and dynamical properties of galaxies.

Our suite of simulations featured different stellar feedback prescriptions {\it Mech} (M), {\it RdTh} (R) and {\it NoFb} (N). These simulations were seeded with a uniform primordial magnetic field of varying comoving strength $B_0$. We also sample each strength with three different orientations for the highest studied redshifts, but only carry out the full suite of simulations to $z = 2$ for one of these orientations, due to both computational costs and a smaller impact than strength variation. All the employed values of $B_0 \sim 10^{-X}$ G ({\it BX}) were chosen to lie below the current observational upper limit ($B_0 < 10^{-9} G$, \citealt{Planck16}): {\it B20}, {\it B14}, {\it B13}, {\it B12}, {\it B11}, and {\it B10}. We studied in detail the manner in which magnetic stresses modify the global properties of galaxies. Finally, we examined how such modifications could be reflected in observable quantities. Our main findings are:
\begin{itemize}
	\item Strong primordial magnetic fields can provide further support against the initial collapse of the galaxy, slightly delaying its formation and temporarily increasing its size both radially and vertically.
	\item After collapse, strong primordial magnetic fields can reduce the radial scale length of the gas disk significantly as the galaxy grows. By redshift $z = 2$, the strongest magnetic field studied ($B \sim 10^{-10}$ G) brings the gas radial scale length down to half the size measured when no significant magnetic fields are present. As a consequence, the stellar disk size is also drastically reduced. Both these reductions are accompanied by a large outward transfer of angular momentum, reflected in the reduction of the spin parameter $\lambda_\text{rot}$.
    \item During the accretion phase, before the disk settles, we observe no clear effects of $B_0$ on the disk scale height. However, once the gas disk has established, magnetic fields in the ISM slightly thicken it. The stellar disk height appears to correlate well with that of the gas disk for moderate values of $B_0$ ($B_0 \sim 10^{-13} - 10^{-11}$ G). However, once $B_0$ becomes stronger ({\it MB10z}), the stellar disk becomes significantly thinner (almost a factor $1/3$). 
	\item Altering the stellar feedback prescription does not induce significant changes on the effects of $B_0$ on morphological properties. However, in the absence of feedback, the effects of magnetic fields on these properties are less prominent until the highest values of $B_0$ are probed ({\it NB11z} and {\it NB10z}). This is likely due to a coupling between stellar feedback and magnetic forces.

    \item Primordial magnetic fields reduce the gas and stellar spin parameters as $B_0$ increases, with this effect becoming especially significant for high primordial magnetic field strengths ({\it B11} and {\it B10}).
    \item No clear effects of $B_0$ on the galaxy's gas turbulence $\sigma_\text{rms}$ are observed. However, a trend is found for $\sigma_\text{rms}^\text{stars}$ to increase as $B_0$ increases and $\lambda_\text{rot}^\text{stars}$ decreases.

	\item Non-negligible magnetic stresses occur for runs with $B_0 > 10^{-13}$ G, which lead to magnetic braking of the galaxies by transporting angular momentum outward. Significant direct spin-down of the galaxy is only found for {\it MB10z}.

    \item In agreement with previous studies, the choice of primordial magnetic field does not alter the total $\Mst$ of the simulated galaxy. However, we find that they influence its SFH. 
    \item Our mock {\it JWST} NIRcam observations display a clear increase of the light concentration parameter $C150W$ (rest frame $V$ band) of the galaxy with $B_0$.
    \item The simulation with the highest primordial magnetic field ({\it MB10z}) produces a galaxy that is redder in rest-frame $U - V$ and $U - J$ colours, as expected for post-starburst systems.
\end{itemize}

Overall, we find that primordial magnetic fields have the potential to influence the growth and properties of galaxies as they evolve. While the effects explored in this manuscript arise from primordial magnetism, which of these remain in place when galactic magnetic fields are seeded through other mechanisms (as e.g. by stellar winds and SNe) remains an open question. Even though our `fiducial' model {\it MB12z} does display changes in the properties of the galaxy, these are generally minor, suggesting that only extreme values of primordial magnetic fields are likely to induce significant modifications.

\section*{Acknowledgements}
The authors kindly thank the referee for their insightful comments and suggestions which have highly improved the quality of this manuscript. This work was supported by the Oxford Hintze Centre for Astrophysical Surveys which is funded through generous support from the Hintze Family Charitable Foundation. This work is part of the Horizon-UK project, which used the DiRAC Complexity system, operated by the University of Leicester IT Services, which forms part of the STFC DiRAC HPC Facility (\href{www.dirac.ac.uk}{www.dirac.ac.uk}). This equipment is funded by BIS National E-Infrastructure capital grant ST/K000373/1 and STFC DiRAC Operations grant ST/K0003259/1. The equipment was funded by BEIS capital funding via STFC capital grants ST/K000373/1 and ST/R002363/1 and STFC DiRAC Operations grant ST/R001014/1. DiRAC is part of the National e-Infrastructure. The authors would like to acknowledge the use of the University of Oxford Advanced Research Computing (ARC) facility in carrying out this work. \href{http://dx.doi.org/10.5281/zenodo.22558}{http://dx.doi.org/10.5281/zenodo.22558}. S.M.A. would like to acknowledge travel support from the Royal Astronomical Society. We thank Clotilde Laigle for insightful comments on using the {\sc SUNSET} software.

\bibliographystyle{mnras}
\bibliography{references}

\appendix

\section{Morphological parameters calculation}
\label{ap:MorphFit}
\begin{figure*}%
\includegraphics[width=1.5\columnwidth]{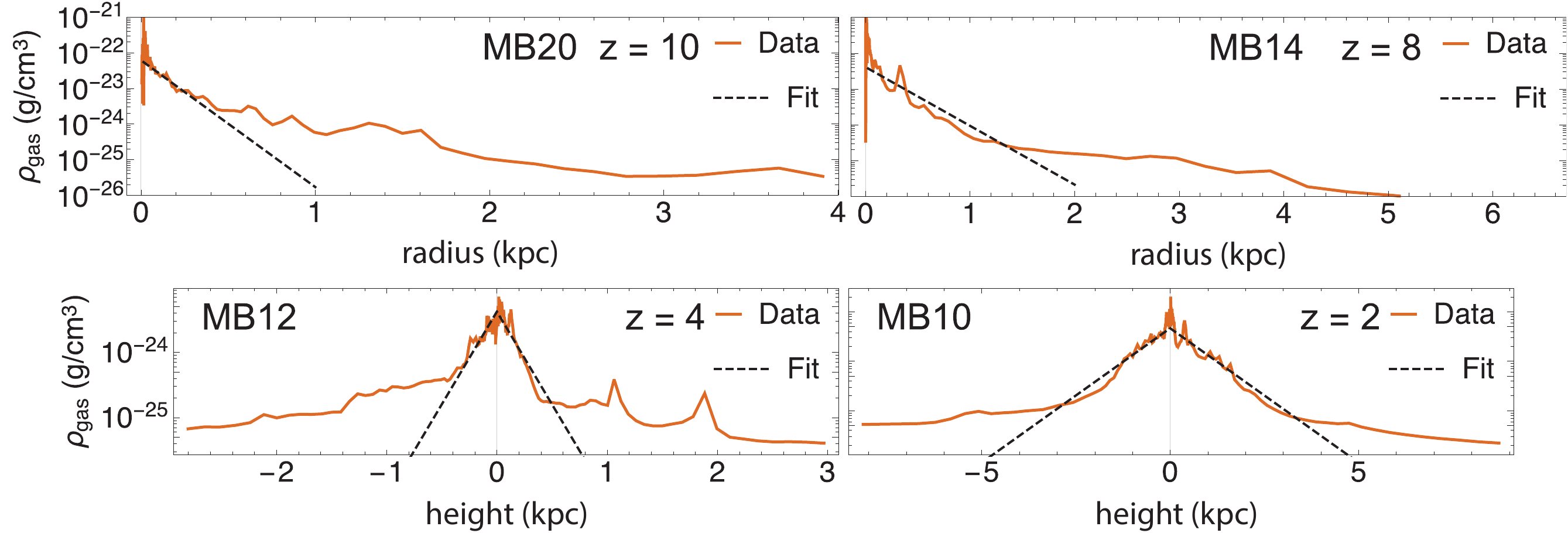}%
\caption{Generic examples of radial (top row) and vertical (bottom row) exponential fits (black dashed lines) to the gas density profiles (orange solid lines) extracted for various outputs of the simulations {\it MB20z}, {\it MB14z}, {\it MB12z}, and {\it MB10z} respectively from top left to bottom right.}%
\label{ExampleFits}%
\end{figure*}

In Section \ref{ss:GalacticPhysics} we make use of the radial scale length $R_s$ and disk scale height $h_s$ of our galaxies to study their morphology. We compute these quantities for each output as follows. $R_s$ is extracted from the radial profile of a cylinder centred on the galaxy (as explained in Section \ref{ss:HaloFinder}), with its vertical axis aligned with the galactic angular momentum. This cylinder has a thickness of $0.1\; r_\text{vir}$, and radially extends out to $0.7\; r_\text{vir}$. To compute $h_s$, we maintain the centering and orientation of the cylinder, but now employ a cylinder with a radial extent of $0.2\; r_\text{vir}$ and a thickness of $0.2\; r_\text{vir}$. We fit an exponential function of the form $f(x) = a \exp{\left(x / b\right)}$ to the resulting density profiles with ($a$, $b$) as free parameters. $x$ is identified either with the radial coordinate $r$ (when $b \rightarrow R_s$) or with the height $h = |z|$ when ($b \rightarrow h_s$). We display some generic examples of $R_s$ and $h_s$ fits in Fig. \ref{ExampleFits}.  

\section{Magneto-thermo-turbulent star formation}
\label{ap:StarForm}

We introduce in this appendix our magneto-thermo-turbulent (MTT) star formation prescription, already described in its thermo-turbulent form by \citet{Trebitsch17, Mitchell18, Rosdahl18}, and to be analysed in more detail in Devriendt et al. (in prep). We present its extension to account for the presence of magnetic fields in simulations \citep[as already employed in][]{KMA18}, but defer the analysis of the effects produced by magnetic fields on the resulting local star formation to future work.

In our simulations, we allow star formation to occur in cells at the highest level of refinement allowed at a given timestep \citep{Rasera06}. The MTT model for star formation accounts for local properties of a grid cell and its environment in order to determine two aspects
\begin{itemize}
    \item whether a gas cell is dominated by the gravity and should continue collapsing beyond the spatial resolution captured by the simulation, and  
    \item what is the star formation efficiency associated with the properties of the gas that this cell and its immediate surroundings contain.
\end{itemize} 
\noindent The first condition serves to restrict star formation in our simulations to regions where the process of collapse should continue, but it cannot proceed due to the limited spatial resolution. To determine whether this is the case for each cell at the highest level of refinement, we define a MTT Jeans length
\begin{equation}
\lambda_\text{J,MTT} = \frac{\pi \sigma_V^2 + \sqrt{36 \pi c_\text{s,eff}^2 G {\Delta\text{x}_\text{cell}}^2 \rho + \pi \sigma_V^4}}{6 G \rho \Delta\text{x}_\text{cell}},
\label{eq:MTTJeans}
\end{equation}
where $G$ corresponds to the gravitational constant, $\rho$ is the gas density, and $\sigma_V$ to the gas turbulent velocity. Following \citep{Federrath12}, we account in this expression for the support of a local small-scale magnetic field against isotropic collapse by defining an effective sound speed
\begin{equation}
    c_\text{s,eff} = c_\text{s} \sqrt{1 + \beta^{-1}},
    \label{eq:effcs}
\end{equation}
where $c_\text{s}$ is the sound speed and $\beta$ is the ratio of thermal to magnetic pressure $\beta = P_\text{thermal} / P_\text{mag} (B)$. Using the modulus of the local magnetic field in the cell, $B = |\vec{B}|$, for the calculation, we neglect anisotropic magnetic forces and instead identify it with a small-scale component that provides isotropic support. When $\Delta\text{x}_\text{cell} > \lambda_\text{J,MTT}$ in a cell, we model the uncaptured collapse through the conversion of a fraction of the gas in the cell into a stellar particle. This is done accounting for a locally defined gas-to-star conversion efficiency parameter: i.e. the star formation efficiency, $\epsilon_\text{ff}$.

Properties of star formation occurring within gas clouds, such as $\epsilon_\text{ff}$, are affected by the characteristics of the hosting clouds \citep{Padoan11, Hennebelle11, Federrath12, Grisdale19}. Amongst them, magnetic fields are important towards determining the star formation efficiency \citep{Hennebelle19}. To account for this variability of star formation efficiency between clounds, we allow  $\epsilon_\text{ff}$ to vary temporally and spatially in our simulations. Therefore, $\epsilon_\text{ff}$ is a local quantity defined for each star forming cell. We convert gas into star particles following a Schmidt law for the star formation rate
\begin{equation}
    \dot{\rho}_\text{star} = \epsilon_\text{ff} \frac{\rho}{t_\text{ff}}.
    \label{eq:SchmidtLaw}
\end{equation}
where we define the free-fall time of the gas $t_\text{ff}$ as
\begin{equation}
    t_\text{ff} = \sqrt{\frac{3\pi}{32 G \rho}}.
    \label{eq:freefall}
\end{equation}
The value of the local star formation efficiency is defined as
\begin{equation}
    \epsilon_\text{ff} = \frac{\epsilon_\text{cts}}{2 \phi_t} \exp{\left(\frac{3}{8} \sigma_s^2\right)} \left[1 + \text{erf}\left(\frac{\sigma_s^2 - s_\text{crit}}{\sqrt{2 \sigma_s^2}}\right) \right],
    \label{eq:LocalEfficiency}
\end{equation}
following the multi-scale PN model from \citet{Padoan11}. In this model, $\sigma_s$ is the dispersion of the logarithm of the gas density to the mean gas density $s = \ln{\left(\rho / \left<\rho\right>\right)}$. The critical density above which post-shock gas in a magnetised cloud is allowed to collapse against magnetic support \citep{Hennebelle11,Padoan11} is defined
\begin{equation}
    s_\text{crit} = \ln{\left(0.067 \; \theta^{-2} \alpha_\text{vir} \mathcal{M}^2 f\left(\beta\right)\right)},
    \label{eq:CritDens}
\end{equation}
\begin{equation}
    f \left(\beta\right) = \frac{\left(1 + 0.925 \beta^{-3/2}\right)^{2/3}}{\left(1 + \beta^{-1}\right)^2},
    \label{eq:BetaFunction}
\end{equation}
with Mach number $\mathcal{M} = \sigma_V / c_\text{s}$. Finally, the virial parameter $\alpha_\text{vir}$ is computed as
\begin{equation}
    \alpha_\text{vir} = \frac{5 \left(\sigma_V^2 + c_\text{s}^2\right)}{\pi \rho G (\Delta\text{x}_\text{cell})^2}
    \label{eq:virialParameter}
\end{equation}
From \citep{Federrath12}, we select for our model $1/\phi_t = 0.57$ and $\theta = 0.33$, corresponding to the best fit values for multi-scale models of star formation in magnetised giant molecular cloud simulations. Equally, we select $\epsilon_\text{cts} = 0.5$, corresponding to the maximum amount of gas that can fall onto stars in the presence of (unresolved) proto-stellar feedback.

Once a cell has been flagged as star forming, it will convert a fraction of its gas into a stellar particle. The probability for a cell to form an integer number of stars $n_*$ follows a Poisson distribution
\begin{equation}
    \mathcal{P}(n_*) = \frac{1}{n_*!}\frac{\mathcal{N}^{n_*}}{e^{\mathcal{N}}}.
\end{equation}
Here, $\mathcal{N}$ is the mean of the distribution
\begin{equation}
    \mathcal{N} = \epsilon_\text{ff} \frac{m_\text{gas}}{m_*} \frac{\Delta t}{t_\text{ff}}
\end{equation}
where $\Delta t$ is the timestep for the level of resolution of the cell, $m_\text{gas}$ is the gas mass in the cell, and $m_*$ is the minimum mass resolution for a star particle ($m_* \sim 2.5 \times 10^3 M_\odot$). The number of stars $n_*$ formed is selected according to the distribution, and determines the mass of the stellar particle ($M_* = n_* m_*$). The described model does not modify the local magnetic energy during star formation. However this will be considered in future work. Finally, to prevent a cell from instantaneously depleting its gas mass, the amount of gas that can be converted into stars in one single timestep is limited to 0.9 $m_\text{gas}$.

\bsp	
\label{lastpage}
\end{document}